\documentclass[a4paper,11pt]{article}
\pdfoutput=1 % if your are submitting a pdflatex (i.e. if you have
\usepackage{jheppub} % for details on the use of the package, please
\usepackage[T1]{fontenc} % if needed
\usepackage{appendix}
\usepackage[dvipsnames]{xcolor}

\usepackage{mathtools,slashed, nccmath, graphicx}
\usepackage{tikz}
\usepackage{subcaption}
\usepackage{graphicx}

\makeatletter
\gdef\@fpheader{\normalfont https://doi.org/10.1007/JHEP10(2024)145}
\makeatother

\title{Bounds on ALP-Mediated Dark Matter Models from Celestial Objects}

\author[a,b]{Tanech Klangburam}
\author[a,b]{Chakrit Pongkitivanichkul}

\affiliation[a]{Khon Kaen Particle Physics and Cosmology Theory Group (KKPaCT),\\ Department of Physics, Faculty of Science, Khon Kaen University, 123 Mitraphap Rd.,\\ Khon Kaen, 40002, Thailand}
\affiliation[b]{National Astronomical Research Institute of Thailand, Chiang Mai 50180, Thailand}

\emailAdd{klangburam.t@gmail.com}
\emailAdd{chakpo@kku.ac.th}

\abstract{We have studied the signals from axion-like particles (ALPs) as dark matter mediators from celestial objects such as neutron stars, brown dwarfs or white dwarfs. We consider the accumulation of dark matter inside the celestial objects using the multiscatter capturing process. The production of ALP from the dark matter annihilation can escape the celestial object and decay into gamma-rays and neutrinos before reaching the Earth. We investigate our model using gamma-ray observations from Fermi and H.E.S.S. and neutrino observations from IceCube and ANTARES. The effective Lagrangian approach allows us to place constraints on the ALP-photon and ALP-fermion couplings. In the gamma-ray channel, our results are able to rule out the existence of ALP with mass up to $\sim \mathcal{O}(10)$ GeV. On the other hand, the neutrino observations can be used to probe a higher mass range with ALP mass up to $\sim \mathcal{O}(100)$ GeV. 
}

\begin{document} 
\maketitle
\flushbottom

\section{Introduction}

After decades of searching for Dark Matter (DM), the models in which DM interacts directly with the Standard Model (SM) particles are strongly disfavoured. In particular, the parameter space of the standard WIMP (Weakly Interacting Massive Particle) models has been severely constrained. This has slowly led to great interest in models with DM interactions via mediators such as Higgs portal model, neutrino portal model, etc \cite{Patt:2006fw,March-Russell:2008lng,Djouadi:2012zc, Falkowski:2009yz,Escudero:2016tzx,Escudero:2016ksa}. Among them, one particular model provides a variety of phenomenological aspects, i.e., a fermion DM with an axion-like particle as the mediator. Axion-like particles (ALPs) are light pseudoscalar particles arising from extensions of the standard model. For example, the String/M theory framework generically provides a plethora of pseudoscalars due to the compactification process of the extra-dimensions \cite{Svrcek:2006yi,Arvanitaki:2009fg}. Since ALPs mainly interact with gauge bosons and fermions via dimension-five couplings, the main phenomenological aspects rely on the ALP-fermion and ALP-photon couplings. One could test the ALP-mediated models using particle colliders such as the beam dump experiments and rare mesons decay. Other interesting tests come from astrophysics such as the neutrino flux coming from supernova SN1987A, effects on stellar evolution of the horizontal branch of the HR-diagram and the number of relativistic degrees of freedom around the time of Big Bang Nucleosynthesis (BBN).
%It has been shown that ALPs could be solutions for the dark matter, dark energy and dark radiation problem \cite{x}.

%Since ALPs mainly interact with gauge bosons through anomaly terms leading to dimension-five couplings with gauge bosons

Recently, there has been interest in DM capturing in celestial objects such as neutron star (NS) \cite{Ilie:2020vec, Nguyen:2022zwb, Bose:2021yhz}, brown dwarf (BD) \cite{Leane:2021ihh, Leane:2020wob} and white dwarf (WD) \cite{Dasgupta:2019juq, Acevedo:2023xnu}. It has been shown that if the multiscattering effect is taken into account, the celestial objects accumulating DM are appealing targets for observation of DM annihilation. The DM-overdense objects could then be observed in 2 different ways depending on the lifetime of the mediator. If the mediator is short-lived and decays inside the celestial object, an extra energy would change the energy budget and the DM can be observed through the lifetime and stability of the celestial objects. On the other hand, if the mediator has a sufficiently long lifetime such that it decays outside the celestial object, the situation allows us to probe DM via indirect detection. Since ALP is weakly interacting and generically has a relatively long lifetime, the indirect detection from the celestial objects could therefore provide an interesting probe for DM with ALP as a mediator.

In this work, we study ALP-mediated DM models using an effective Lagrangian approach. We then use the DM-nucleon cross-section to calculate the multiscatter capture rate of the celestial objects in the DM-rich region. 
%\textcolor{red}{We are interested in the assumption that the ``mediator from DM annihilation is ALPs," ALP is weakly interacting such that its decay width is typically longer than the size of celestial objects. Therefore, the indirect detection probes of the model using the celestial objects will be the focus of our study.}
%We are interested in the case where ALP produced from annihilation has a sufficiently long lifetime such that the ALP decays into SM particles \textcolor{red}{outside the celestial object.} 
Using the distribution of celestial objects near the galactic center and the DM generalized Navarro-Frenk-White (NFW) density profile, we are able to produce gamma-ray fluxes and neutrino fluxes from the DM annihilation in this model. The gamma-ray fluxes from NSs, BDs and WDs can be tested against Fermi satellite and H.E.S.S. data, as pointed out in \cite{Leane:2021ihh} (NSs and BDs) and \cite{Acevedo:2023xnu} (WDs).
%The gamma-ray fluxes from NSs, BDs and WDs can be tested against Fermi satellite and H.E.S.S. data. 
The constraints from gamma-ray observation using NSs, BDs and WDs as targets are able to rule out a significant portion of parameter space of ALP couplings and ALP mass.
The fluxes from neutrino produced from DM annihilation inside BDs and WDs could be detected via IceCube experiment and ANTARES neutrino telescope. %, whereas in the NS case, the neutrino fluxes are currently not observable by the neutrino observations. 
We show that the bounds from neutrino fluxes are also stronger than existing constraints from various experiments and observations. In our analysis, we are able to provide constraints up to $\sim\mathcal{O}(100)$ GeV in ALP mass, which emphasis the importance of future probes on gamma-ray and neutrino observations.

%, whereas in the NS case, the neutrino fluxes are too low for any current observations. 
%The constraints from gamma-ray fluxes perform better than the ones from neutrino fluxes due to the difference in the observed fluxes from the experiments.

%because the observed fluxed from neutrino is one order of magnitudes higher than that of the photon. 
%The constraints from neutrino using BDs as the targets are able to cover the parameters space from other constraints in higher ALP mass.

%The annihilation of DM inside the compact object can produce the ALP that decays into SM particles. 

%For the case that ALP decays outside of the compact objects, the constraints on mass and axion-photon couplings can be drawn. We then use the gamma-ray observation such as Fermi and H.E.S.S to obtain the limits. We found that the 

%Now the ALP with long enough lifetime is in the range of other experiments where we are interested in and the shorter lifetime case will be studied in the upcoming paper.

%In this paper, we are considering the multiscatter capture in the context of ALPs mediated dark matter models. The 

% Axion like particle \\
% Axion mediated dark matter models \\
% Dark matter multiscatter capturing \\
% A good review can be found in \cite{Bauer:2017qwy} \\

\section{ALP-mediated Dark Matter Simplified Model}

\begin{figure}[h]
\centering
    
    \begin{subfigure}{0.2\textwidth}
        \includegraphics[width=\textwidth]{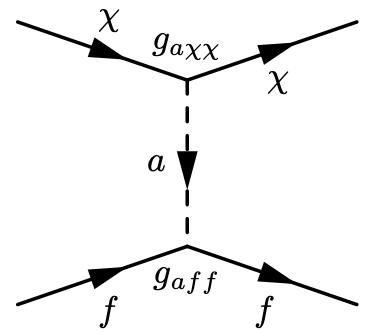}
        \caption{\footnotesize $\chi f \rightarrow \chi f $}
        \label{fig:xxff}
    \end{subfigure}
    \hspace*{.20\textwidth}
    \begin{subfigure}{0.2\textwidth}
        \includegraphics[width=\textwidth]{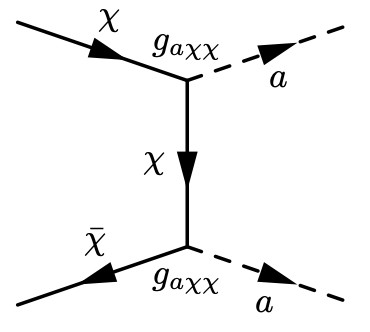}
        \caption{\footnotesize $\chi\bar{\chi}\rightarrow a a$}
        \label{fig:xxaa}
    \end{subfigure}
    
    \begin{subfigure}{0.2\textwidth}
        \includegraphics[width=\textwidth]{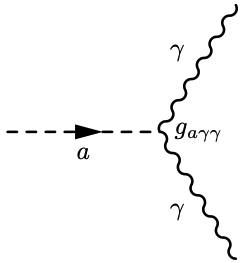}
        \caption{\footnotesize $a \rightarrow \gamma\gamma$}
        \label{fig:gagg}
    \end{subfigure}
    \hspace*{.20\textwidth}
    \begin{subfigure}{0.2\textwidth}
        \includegraphics[width=\textwidth]{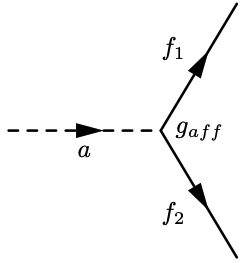}
        \caption{\footnotesize $a\rightarrow f f$}
        \label{fig:gaff}
    \end{subfigure}
    \caption{Panel (a) shows the diagram of a non-relativistic scattering of DM with the SM and panel (b) shows the diagram of DM annihilation into ALP. Panel (c) shows the diagram of ALP effectively decays into gamma-rays and the diagram of ALP effectively decays into SM particles is shown in panel (d).}
    %\caption{\textcolor{red}{Diagrams a non-relativistic scattering of dark matter with the standard model where the ALP propagates an interaction (left), an ALP effectively decays into gamma-rays (middle) and an ALP effectively decays into SM particles (right).}}
    \label{fig:diagram}
\end{figure}

In this model, the SM is extended by a Dirac fermion, $\chi$, and a pseudoscalar ALP, $a$. The effective Lagrangian is given by
\begin{eqnarray}
    \mathcal{L}\supset \frac{1}{2}\partial_\mu a \partial^\mu a - \frac{1}{2}m^2_a a^2 + \sum_f \frac{m_f}{f_a}C_f\bar{f}i\gamma_5 f a + \frac{m_\chi}{f_a}C_\chi\bar{\chi} i \gamma_5 \chi a  - \frac{g_{a\gamma\gamma}}{4}F_{
    \mu\nu}\Tilde{F}^{\mu\nu}a, \label{eq:L}
\end{eqnarray}
where $f$ is any SM fermion with mass $m_f$, the masses of ALP and DM are $m_a$ and $\ m_\chi$, respectively. $F_{\mu\nu} = \partial_\mu A_\nu - \partial_\nu A_\mu$ is the $U(1)_{\rm EM}$ field strength tensor ($\Tilde{F}^{\mu\nu} = \frac{1}{2} \epsilon^{\mu\nu\alpha\beta}F_{\alpha\beta}$) and $g_{a\gamma\gamma}$ is an effective ALP-photon coupling. For convenience, we define the effective ALP couplings by 
\begin{eqnarray}
    g_{aff} = m_f C_f/f_a\quad \text{and}\quad g_{a\chi\chi} = m_\chi C_\chi/f_a, \label{eq:c} 
\end{eqnarray}
where the ALP-fermion couplings $C_f$ are assumed to be universal for any fermion, i.e., $C_f$ and $g_{aff} \propto m_f$. Instead of using a comprehensive analysis on the effective Lagrangian as shown in \cite{Bauer:2017ris}, we are using a simplified approach to the ALP couplings such that all loop corrections are already included in $g_{a\gamma\gamma}$ and $g_{aff}$. The Dirac fermion $\chi$ is assumed to be the DM particle. The ALP, $a$, acts as the mediator between SM and the hidden sector where $g_{aff}$ acts as the connector coupling and $g_{a\chi\chi}$ acts as the hidden sector coupling. 
%From the Lagrangian, we can interpret that the DM can interact with SM particles through the mediator which is an ALP. The ALP can interact with the SM and DM and ALP can also decay into fermions or photons. 
Figure \ref{fig:diagram} illustrates the diagrams involved in this work. The diagram of DM-nucleon scattering is shown in figure \ref{fig:xxff} and the DM annihilation is depicted in figure \ref{fig:xxaa}. The diagrams of ALP decays into 2 photons and 2 SM particles are shown in figure \ref{fig:gagg} and \ref{fig:gaff}, respectively.
%Figure \ref{fig:diagram} shows the diagram of the scattering between DM and nucleons which is mediated by ALP and the diagrams of ALP decays into 2 photons and 2 SM particles, % (specifically neutrinos if kinematically allowed) respectively. 
The different regimes of interactions were studied for the freeze-in/out scenarios in \cite{Hochberg:2018rjs,Bharucha:2022lty, Ghosh:2023tyz, Dror:2023fyd,Armando:2023zwz}, therefore we will not focus on the DM production mechanism in this paper.
%The mechanisms that ALP is a mediator between DM and SM were studied in various models []. 

%\begin{figure}[h]
%\centering
%    \includegraphics[width=4cm]{image/diagram.png}
%    \includegraphics[width=4cm]{image/diagram-2a.png}
    %\hspace{1cm}
%    \includegraphics[width=3cm]{image/gagg.png}
    %\hspace{1cm}
%    \includegraphics[width=3cm]{image/gaff.png}
%    \caption{Diagrams a non-relativistic scattering of dark matter with the standard model where the ALP propagates an interaction (left), an ALP effectively decays into gamma-rays (middle) and an ALP effectively decays into SM particles (right).}
%    \label{fig:diagram}
%\end{figure}

As DM condenses around the celestial objects in the DM-rich region, the DM can fall into the objects by the multiscatter capture \cite{Bramante:2017xlb, Ilie:2020vec}. We will discuss this mechanism in the next section. DM can lose its kinetic energy by scatters with the components of the object repeatedly and then falling into the celestial object.
%, we will investigate the productions that came from the DM annihilation of the DM that accumulates inside the object. 
The scattering cross-section of $\chi f \rightarrow \chi f$ in the low-energy limit is
\begin{eqnarray}
    \sigma(\chi f \rightarrow \chi f) = \frac{g_{aff}^2 g_{a\chi\chi}^2}{2\pi{m_a^4}} \frac{m^4_\chi+ m^4_f + m^2_\chi m^2_f}{(m_\chi + m_f)^2}. \label{eq:xsec}
\end{eqnarray}
%Each time that DM scatters with the components of the object, 
%DM will lose kinetic energy and move toward the center of the object. 
DM will then accumulate and start to annihilate inside the object producing a long-lived mediator which is an ALP via the process $\chi\bar{\chi} \rightarrow aa$. Due to the difference in mass between DM and ALP, the ALP will move relativistically out of the target and eventually decay into SM particles before reaching the earth. The signals from these particles can be observed via gamma-ray telescopes or neutrino observatory which will be discussed in later sections.
%\textcolor{red}{In lower ALP mass, the final state can be predicted to consist of photons, neutrinos or charged leptons. The charged leptons can be affected by the galactic magnetic field, which makes it difficult to determine the directions of the signals. Consequently, we will focus only on gamma-rays and neutrinos.}
The decay width of ALP to gamma-rays is given by
\begin{eqnarray}
    %\tau^{-1}_{a\rightarrow \gamma\gamma} = 
    \Gamma_{a\rightarrow \gamma\gamma} = \frac{1}{64\pi}g^2_{a\gamma\gamma}m_a^3, \label{eq:lifetime}
\end{eqnarray}
and the decay width of ALP to a pair of fermions is written as
\begin{eqnarray}
    %\tau^{-1}_{a\rightarrow ff} = 
    \Gamma_{a\rightarrow ff} = \frac{g_{aff}^2n_f^c m_a}{8\pi}\sqrt{1-\frac{4m^2_f}{m_a^2}}. \label{eq:lifetime_nu}
\end{eqnarray}
Note that in the low mass range of ALP, other gauge boson channels are not kinematically allowed.
%The detailed analysis of the effective Lagrangian approach can be found in \cite{Bauer:2017ris}. }
%Note that other gauge bosons channel are generally loop suppressed \cite{Bauer:2017ris} 
In order to combine constraints from various experiments, we instead choose to parameterize the coupling $g_{a\gamma\gamma}$ with the branching ratio parameter $B$, i.e., 
%In this work, we assume that the branching ratio of ALP decays into gamma-rays is given by the parameter $B$, i.e., 
\begin{eqnarray}
    \text{BR}(a\rightarrow\gamma\gamma) = \frac{\Gamma_{a\rightarrow\gamma\gamma}}{\Gamma_\text{tot}} \equiv B, \quad \Gamma_{\text{tot}} = \Gamma_{a\rightarrow\gamma\gamma} + \sum_{f} \Gamma_{a\rightarrow ff}. \label{eq:BRg}
\end{eqnarray}
%\quad \text{and}\quad \text{BR}(a\rightarrow ff) = \frac{\Gamma_{a\rightarrow ff}}{\Gamma_\text{tot}} = 1-B, \label{eq:BR}
%where $B$ is less than 1, and in this work, we will assume $B=0.1$ for the rest of the calculations. 
%Therefore, $g_{aff}$ can be recast to $g_{a\gamma\gamma}$ as
Therefore, $g_{a\gamma\gamma}$ can be recast to $g_{aff}$ as
\begin{eqnarray}
    g_{aff}^2 = \frac{1-B}{B} \sum_f \frac{m_a^2}{8n_f^c}\frac{1}{\sqrt{1-\frac{4m_f^2}{m_a^2}}}\times g_{a\gamma\gamma}^2,  \label{eq:relate}
\end{eqnarray}
where the universality of $g_{aff}$ is used and the summation is made over the kinematically allowed fermions.
%\textcolor{blue}{CP: How about we move the discussion on ALP trapping to later section? Like section 5 on Detection?}
One might also concern about the possibility of ALP being trapped inside the object before escaping, due to ALP interaction with the components of the object such as nuclei. However, we argue in appendix \ref{sect:app1} that the interaction length of ALP-nucleon interaction is much longer than a typical radius of the objects that we are interested in.

%The probes on the ALP coupling have been studied in various methods (see \cite{Dolan:2017osp, Bharucha:2022lty} for a review). For example, the astrophysical sources such as SN 1987A can provide us the constraints on ALP coupling that $g_{a\gamma\gamma}<6\times10^{-9}$ GeV$^{-1}$ for small ALP masses \cite{Jaeckel:2017tud}. The red giant branch in several clusters \cite{Straniero:2020iyi} give the upper bounds on the ALP coupled to electrons as $g_{aee}< 1.48\times10^{-13}$, which is also similar to the analysis of the red giant in the Galactic globular cluster $\omega$ Centauri $g_{aee}< 1.3\times10^{-13}$ \cite{Capozzi:2020cbu}. The ALP coupled to both SM and DM has been recently studied in the context of thermal history \cite{Bharucha:2022lty, Balazs:2022tjl, Ghosh:2023tyz, Dror:2023fyd, Leane:2023woh}.

We are interested in the case that ALPs were generated by the DM annihilation from the celestial objects. We primarily aimed to constrain the ALP-fermion coupling, $g_{aff}$ from the gamma-rays and neutrinos observations. The celestial objects such as NSs, BDs and WDs are good targets since they are densely distributed around the galactic center where the DM density is generically higher.
%known as the DM-rich region.

\section{Dark Matter Multiscatter Capture}

%As the celestial bodies move through the DM halo, 
DM from the Galactic halo can start to fall into the celestial object if the DM is sped up to the escape velocity of the object at the surface of the celestial body which is 
\begin{eqnarray}
    v_\text{esc} = \sqrt{G_N M/R},
\end{eqnarray}
where $G_N$ is the gravitational constant, $M$ and $R$ are the total mass and the radius of the celestial body, respectively.
%where $G_N$ is the gravitational constant, $M$ and $R$ are the total mass and the radius of the object. 
%\textcolor{red}{The large gravitational field at the surface of the celestial objects can cause the blue shifting of DM of $\chi\approx 1 - \sqrt{1-2G_N M/R}$, this gives $v_\text{esc}\simeq \sqrt{2\chi}$}. 
As DM particles transit through the celestial body, they can scatter with stellar materials and lose their kinetic energy. Once the velocity of DM drops below the escape velocity of the celestial objects, DM is captured. The capturing process can occur via single or multiple scatters \cite{Kouvaris:2010vv,Bramante:2017xlb,Dasgupta:2019juq,Ilie:2020vec,Ilie:2021iyh,Leane:2021ihh}. The probability for a given DM to undergo $N$ scatter is given by 
\begin{eqnarray}
    p_N(\tau) = 2\int_0^1 dy \frac{y e^{y\tau} (y\tau)^N}{N!},
\end{eqnarray}
where $\tau = \frac{3}{2}\frac{\sigma_{\chi n}}{\sigma_\text{sat}}$ is the optical dept. The saturate cross-section is defined as $\sigma_\text{sat} \equiv \pi R^2/N_n,$ where $N_n$ is the number of nucleons inside the celestial object. The meaning of saturation cross-section $\sigma_{sat}$ is that the DM has a mean free path of the size of the object, such that, on average, the DM will scatter once at this cross-section. The probability can be approximated in the limit of single scatter $(\tau \leq 1)$ and multiple scatter $(\tau \gg 1)$ \cite{Ilie:2020vec} as  
\begin{eqnarray}
    p_N(\tau) \approx 
    \begin{dcases}
    \frac{2\tau^N}{N!(N+2)} + \mathcal{O}(\tau^{N+1}), & \text{if }\quad \tau \leq 1\\
    \frac{2}{\tau^2}(N+1)\Theta(\tau-N). & \text{if }\quad \tau \gg 1
    \end{dcases}
\end{eqnarray}

The capture rate after DM scattered $N$ times and became trapped in the celestial object is given by \cite{Leane:2022hkk}
\begin{eqnarray}
    C_N(\tau) &=& f_\text{cap} \times \pi R^2 p_N(\tau) \frac{\sqrt{6}n_\chi}{3\sqrt{\pi}\bar{v}}  \nonumber \\
    & & \times \left[ (2\bar{v}^2 + 3v^2_{esc}) - (2\bar{v}^2 + 3v^2_N)\ \text{exp}\left( - \frac{3}{2}\frac{(v_N^2 - v^2_{esc}) }{\bar{v}^2} \right)  \right], \label{eq:capturerate}
\end{eqnarray}
where $n_\chi = \rho_\chi(r)/m_\chi$ is the local number density of DM and $\bar{v}$ is the DM dispersion velocity.
After scattering $N$ times, the typical velocity of DM which takes into account the energy loss in each scattering is
\begin{eqnarray}
    v_N = v_\text{esc}(1 - \langle z \rangle \beta)^{N/2},
\end{eqnarray}
with $\beta = 4m_\chi m_n/(m_\chi + m_n)^2$ where $m_n$ is the mass of the target particles in the star and $z$ is related to the scattering angle, $z \in [0,1]$. 
In the regime of DM mass is much less than $m_n$, $m_\chi\ll m_n$, and $v_\text{esc} < \bar{v}$, the reflection factor is given by \cite{Leane:2023woh, Neufeld:2018slx}
\begin{eqnarray}
    f_\text{cap}\approx \frac{2}{\sqrt{\pi N_\text{scatter}}}=\left[ \frac{4}{\pi} \frac{\log(1-\langle z \rangle \beta)}{\log(v_\text{esc}^2/\bar{v}^2)} \right]^{1/2},
\end{eqnarray}
where $N_\text{scatter}$ is the total number of scatters required for capture. The factor $f_\text{cap}$ is taken into account the back-scattering effect of DM when passing through the star and get captured. For $m_\chi > m_n$ and $v_\text{esc} > \bar{v}$, the reflection factor is $f_\text{cap}\approx 1$.
%In the regime of DM mass is comparable with $m_n$ and $v_\text{esc} < \bar{v}$, the reflection factor is given by \cite{Neufeld:2018slx}
%\begin{eqnarray}
%    f_\text{cap}\approx \frac{2}{\sqrt{\pi N_\text{scatter}}}=\left[ \frac{4}{\pi} \frac{\log(1-\langle z \rangle \beta)}{\log(v_\text{esc}^2/\bar{v}^2)} \right]^{1/2},
%\end{eqnarray}
%where $N_\text{scatter}$ is the total number of scatters required for capture. The factor $f_\text{cap}$ is taken into account the back-scattering effect of DM when passing through the star and get captured. For $m_\chi > m_n$ and $v_\text{esc} > \bar{v}$, the reflection factor is $f_\text{cap}\approx 1$.}

The total capture rate for a single celestial body is then given by
\begin{eqnarray}
    C = \sum_{N=1}^\infty C_N. \label{eq:csum}
\end{eqnarray}
For sufficiently large $N$, $2\bar{v}^2 + 3v^2_N$ becomes much larger than $\bar{v}_N^2$, therefore, the exponential term in eq.~(\ref{eq:capturerate}) can be neglected, the maximum capture rate is given by
\begin{eqnarray}
    C_\text{max} = f_\text{cap}\times \pi R^2 n_\chi(r)v_0 \left( 1 + \frac{3}{2}\frac{v^2_\text{esc}}{\bar{v}(r)^2} \right), \label{eq:Cmax}
\end{eqnarray}
%where the geometric capture rate is
%\begin{eqnarray}
    %C_\text{geo} = \pi R^2 n_\chi(r)v_0 \left( 1 + \frac{3}{2}\frac{v^2_\text{esc}}{\bar{v}(r)^2} \right), \label{eq:Cgeo}
    %\xi(v_p, \bar{v}(r)), 
%\end{eqnarray}
where $v_0 = \sqrt{8/3\pi}\bar{v}$. We calculate the capture rate by using the \texttt{Asteria} package \cite{Leane:2023woh} where they have provided the treatment on $f_\text{cap}$ and $p_N(\tau)$ for calculating the capture rate. We compute the scattering cross-section from eq.~(\ref{eq:xsec}) to use as the input to the \texttt{Asteria} package along with the properties of DM particles and celestial objects.

\begin{figure}[h]
\centering
    \includegraphics[height=6.5cm]{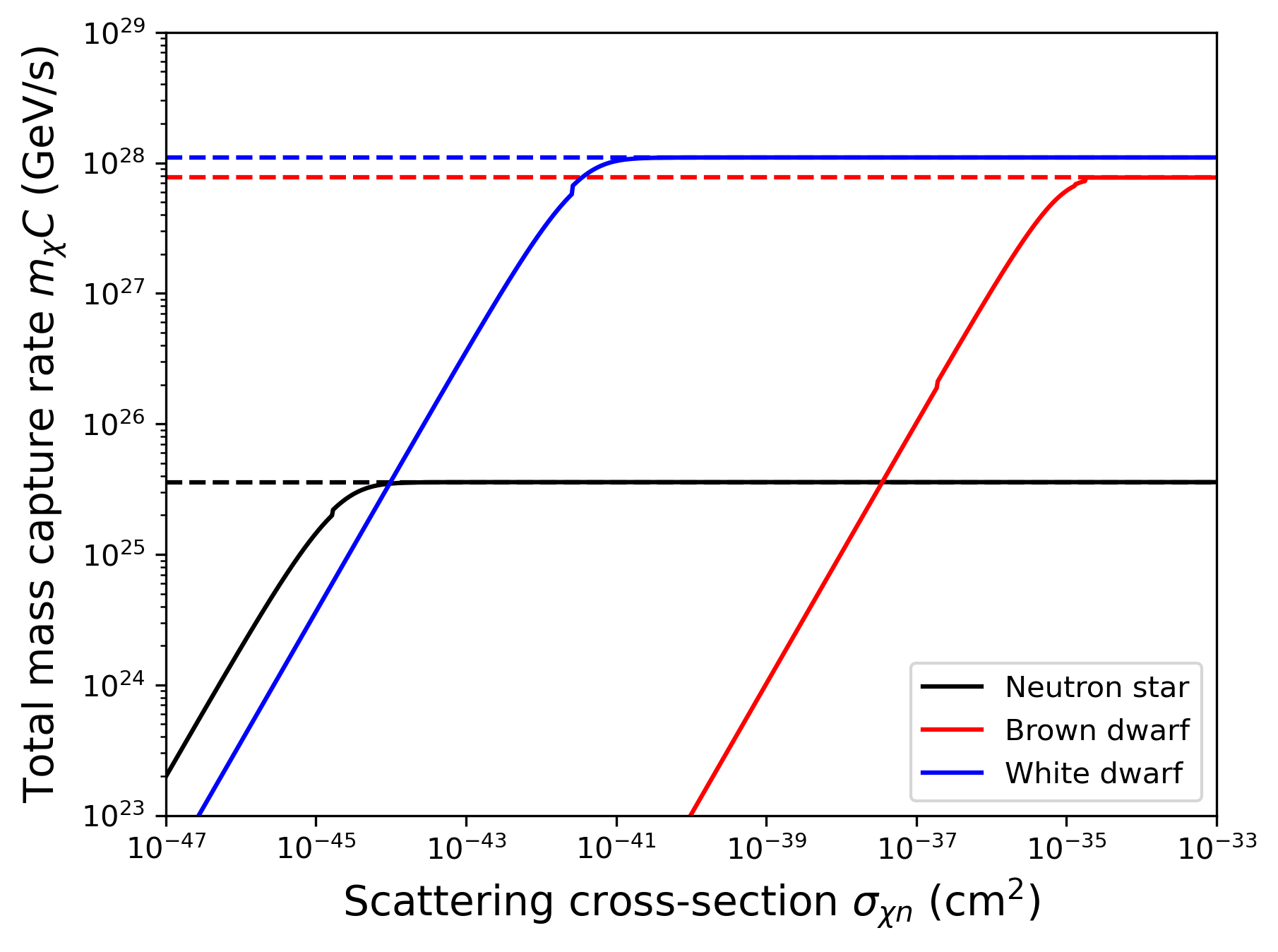}
    \caption{The mass capture rate $m_\chi C$ as a function of DM-nucleon scattering cross-section $\sigma_{\chi n}$ for a NS, BD and WD. The dashed line indicates $m_\chi C_\text{max}$. The DM model is applied with $m_\chi = 10^3$ GeV, $\rho_\chi = 0.42$ GeV/cm$^3$ and $\bar{v} = 220$ km/s.}
    \label{fig:Cmax}
\end{figure}

Figure \ref{fig:Cmax} shows the mass capture rate for a NS (black), BD (red) and WD (blue) as a function of DM-nucleon scattering cross-section $\sigma_{\chi n}$. As $\sigma_{\chi n}$ increases, the number of scatters $N$ also increases, once it goes above the certain number of scatters that is sufficiently large enough to capture DM, the capture rate in eq.~(\ref{eq:capturerate}) (solid line) is approximated as the maximum capture rate (dashed line) described by eq.~(\ref{eq:Cmax}). 
It can be seen that the maximum capture rate of WD is slightly higher than that of BD, and both objects are 100 times larger than the NS maximum capture rate. This is because the effective radius of WD/BD is larger than the radius of NS, which causes more DM flux to pass through the object. Note that, BD requires a larger DM-nucleon scattering cross-section compared to WD in order to reach the maximum capture rate. Despite its larger size, BD has less mass than WD. Consequently, WD is denser than BD which requires a small DM-nucleon scattering cross-section to reach the maximum rate.

%It can be seen that the BD capture rate is 100 times larger than the NS capture rate. This is because the effective radius of BD is much larger than the radius of NS, which causes more DM flux to pass through the object.

%\begin{figure}[h]
%\centering
%
%\begin{subfigure}{0.48\textwidth}
%    \includegraphics[width=\textwidth]{image/max_total_NS.png}
%    \caption{Neutron star.}
%    \label{fig:cmaxns}
%\end{subfigure}
%\hfill
%\begin{subfigure}{0.48\textwidth}
%    \includegraphics[width=\textwidth]{image/max_total_BD.png}
%    \caption{Brown dwarf.}
%   \label{fig:cmaxbd}
%\end{subfigure}
%
%\begin{subfigure}{0.48\textwidth}
%    \includegraphics[width=\textwidth]{image/max_total_WD.png}
%    \caption{White dwarf.}
%    \label{fig:cmaxwd}
%\end{subfigure}
%\caption{The mass capture rate $m_\chi C$ as a function of nucleon scattering cross-section $\sigma_{\chi n}$ for \textcolor{red}{a NS, BD and WD}. The red dash line indicates $m_\chi C_\text{max}$. The DM model is applied with $m_\chi = 10^3$ GeV, $\rho_\chi = 0.42$ GeV/cm$^3$ and $\bar{v} = 220$ km/s.}
%\label{fig:Cmax}
%\end{figure}

In order to calculate the total capture rate for all celestial objects within a given system (e.g. the galactic center or galaxy clusters), we need to take into account the number density of the objects within radii $r_1$ and $r_2$, the total capture rate is \cite{Leane:2021ihh}
\begin{eqnarray}
    C_\text{total} = 4\pi\int_{r_1}^{r_2} r^2 n_\star (r) C \ dr, \label{eq:ctotal}
\end{eqnarray}
where $n_\star (r)$ is the number density of the celestial objects and $C$ is the capture rate of a single object from eq.~(\ref{eq:capturerate}). The number density of the celestial objects will be described in the following section.

%To calculate the DM capture rate, we use \texttt{Asteria} [2309.00669]. 

\section{The Galactic Center}

In this section, we introduce the specific model of the DM velocity distribution and the target number density. The DM velocity dispersion and DM density affect the rate at which DM would fall into the gravitational potential and eventually fall into the object. 
%The dispersion velocity $\bar{v}$ can determine the energy lost in the scattering process, which plays an important role in DM capture. 
At a given radius $r$ from the galactic center, the dispersion velocity is %$ \bar{v} = \sqrt{3/2} v_c(r),$ 
\begin{eqnarray}
    \bar{v} = \sqrt{3/2} v_c(r), 
\end{eqnarray}
where $v_c$ is the circular velocity at radius $r$, which depends on the total mass of the sphere of radius $r$: %$v_c = \sqrt{{G_N M(r)}/{r}}$.
\begin{eqnarray}
    v_c = \sqrt{{G_N M(r)}/{r}}.
\end{eqnarray}
We use the model for the mass distribution following ref.~\cite{Sofue:2013kja}. This model has five components which are the central supermassive black holes $M_\text{BH} = 4\times10^6$ M$_\odot$, inner and outer bulges $(\rho_\text{inner}$ and $\rho_\text{outer})$, a disk component $(\rho_\text{disk})$ and DM halo $(\rho_\text{DM})$. Combining all five components, the total mass is
\begin{eqnarray}
    M(r) = M_\text{BH} + 4\pi \int_0^r r^2 (\rho_\text{inner} + \rho_\text{outer} + \rho_\text{disk} + \rho_\text{DM})\ dr.
\end{eqnarray}
The DM generalized Navarro-Frenk-White (NFW) density profile \cite{Navarro:1995iw} is
\begin{eqnarray}
    \rho_\text{DM} = \frac{\rho_0}{(r/r_s)^\gamma (1 + (r/r_s)^{1-\gamma})}, \label{eq:rho_DM}
\end{eqnarray}
where $\rho_0 = 0.42$ GeV/cm$^3$ is the local DM density, the scale radius is $r_s = 12$ kpc and the slope index is ranging from $\gamma = 1-1.5$. The model of the inner bulge, outer bulge and disk is assumed to be an exponential sphere model:
\begin{eqnarray}
    \rho_i = \rho_{0,i} e^{-r/a_i}, \label{eq:rhomodel}
\end{eqnarray}
where the parameters of each component are shown in table \ref{table:1}.

\begin{table}[h]
\centering
\begin{tabular}{|c|c|c|c|}
    \hline
     Mass component & Total mass (M$_\odot$) & Scale radius (kpc) & Center density (M$_\odot$ pc$^{-3}$) \\ \hline
     Inner bulge & $5\times10^7$ & 0.0038 & $3.6\times10^4$ \\
     Outer bulge & $8.4\times10^9$ & 0.12 & $1.9\times10^2$ \\
     Disk & $4.4\times10^{10}$ & 3.0 & 15 \\
    \hline
\end{tabular}
\caption{The parameters in eq.~(\ref{eq:rhomodel}) for describing the mass distribution of each component of the galactic center, from \cite{Sofue:2013kja}.}
\label{table:1}
\end{table}

Neutron star (NS) is the collapsed core of a massive star ($10-25$ $M_\odot$), NS is composed almost entirely of degenerate neutrons with a mass ranging from $1-1.5$ M$_\odot$ and a radius $R_\text{NS}\simeq10$ km. The number densities of NSs and black holes in regions around the galactic center have been numerically studied with the nuclear star cluster dynamics \cite{Generozov:2018niv}. Two types of cluster models were described, `Fiducial' and `Fiducial $\times$ 10', both of which are potentially good candidates for an NS distribution in the nuclear star clusters. The NS radial number density distribution from the `Fiducial $\times$ 10' model is given by \cite{Leane:2021ihh}
\begin{eqnarray}
    n_\text{NS}(r) =   
    \begin{dcases}
    5.98\times10^3 \left( \frac{r}{1\ \text{pc}}\right)^{-1.7} & \text{if }\quad 0.1\ \text{pc}\ <r <2\ \text{pc}\\
    2.08\times10^4 \left( \frac{r}{1\ \text{pc}}\right)^{-3.5} & \text{if }\quad r >2\ \text{pc}
    \end{dcases}. \label{eq:NS}
\end{eqnarray}

On the other hand, Brown dwarf (BD) is a failed star that is not massive enough to process the nuclear fusion of hydrogen into helium in its core. A huge amount of BDs are expected to be presented in our galaxy. The Milky Way may contain $25-100$ billion BDs \cite{10.1093/mnras/stx1906}. To obtain the radial distribution function of BDs, we follow the works in refs. \cite{Kroupa:2011aa,Amaro-Seoane:2019umn}. They extended the Kroupa Initial Mass Function (IMF) into the BD IMF which is described by the broken power law function, $\frac{dN_\text{BD}}{dm} \propto m^{-\alpha}$, where $\alpha=0.3$, $N_\text{BD}$ is the number of BD and $m$ is the mass of BD. The number density of the BDs with the mass range from $0.01-0.07$ M$_\odot$ is given by \cite{Leane:2021ihh}
\begin{eqnarray}
    n_\text{BD}(r) = 7.5\times10^4\left( \frac{r}{1\ \text{pc}}\right)^{-1.5}. \label{eq:BD}
\end{eqnarray}
In our calculation, %we will use the average mass of BDs as $M_\text{BD} = 0.0378$ M$_\odot$ and their radius is assumed to be equal to the Jupiter radius $R_\text{BD} = R_\text{J} = 69,911$ km. 
the average mass of BDs is taken to be $M_\text{BD} = 0.0378$ M$_\odot$ and the radius is assumed to be equal to the Jupiter radius $R_\text{BD} = R_\text{J} = 69,911$ km.

Another target is a White dwarf (WD), which is considered as the endpoint of the stellar evolution for stars that do not have enough mass to become neutron stars or black holes. To obtain the number density of WDs, we follow the work in ref.~\cite{Acevedo:2023xnu} where they parameterized the number density distribution of WDs as the power law with slope as
\begin{eqnarray}
    n_\text{WD}(r) = n_\text{WD}(r_0)\left(\frac{r}{r_0}\right)^{-\alpha}, \label{eq:WD}
\end{eqnarray}
where $\alpha = 1.4$ is adopted from the analytical estimate in ref.~\cite{Alexander:2008tq}. The normalization factor $n_\text{WD}(r_0)$ were estimated to be $5.79\times10^{5}$ pc$^{-3}$ with $r_0 = 1.5$ pc and this factor valid between $10^{-4}$ pc and $1.5$ pc. In \cite{Bell:2021fye}, the WD core is modeled, providing benchmarks for the properties of 100\% carbon-12 WDs. In this work, we will use one of the benchmarks where $M_\text{WD} = 0.49$ $M_\odot$ and $R_\text{WD} = 9,390$ km.

\section{Detection}

Once DM becomes trapped in the celestial bodies (NS, BD or WD), it has two possible fates. First, if DM does not self-annihilate, its density will rise near the core of the celestial object and it can lead to black hole formation and collapse. The second case is that the DM is annihilated with each other inside the celestial body. The evolution of the number of DM particles \cite{Kouvaris:2010vv} inside an object $N(t)$ is governed by an interplay between DM capture rate and DM annihilation rate, i.e.,
\begin{eqnarray}
    \frac{d N(t)}{dt} = C_\text{total} - C_A N(t)^2, \label{eq:eom}
\end{eqnarray}
where $C_\text{total}$ is the total capture rate from eq.~(\ref{eq:ctotal}) and $C_A = \langle \sigma_A v \rangle/V_\text{eff}$ is the average thermal annihilation cross-section over the effective volume of the celestial object body $V_\text{eff} = 4\pi R^3/3$. The solution of eq.~(\ref{eq:eom}) reads
\begin{eqnarray}
    N(t) = \sqrt{\frac{C_\text{total}}{C_A}}\tanh\left(\frac{t}{t_\text{eq}}\right),
\end{eqnarray}
where $t_\text{eq} = 1/\sqrt{C_A C_\text{total}}$ is the time scale required for the celestial object body to reach the DM equilibrium between capture and annihilation. 
%For the case of a small capture rate, the equilibrium might not reach within the lifetime of the celestial bodies. However the fluxes are still detectable, the assumption of annihilation rate for the case of non-self-conjugate DM particles at time $t$ is
%\begin{eqnarray}
%    \Gamma_\text{ann} = \frac{N(t)^2}{4 V_{eff}}\langle \sigma_A v \rangle.
%\end{eqnarray}
Once it reaches the equilibrium, $t>t_\text{eq}$, the number of DM becomes time-independent, and the annihilation rate $\Gamma_\text{ann}$ is simply
\begin{eqnarray}
    \Gamma_\text{ann} = \frac{C_\text{total}}{2},
\end{eqnarray}
where the factor $1/2$ comes from the fact that the DM annihilation process involves 2 DM particles. Since it has been shown that within the lifetime of the universe, the equilibrium between DM capturing rate and DM annihilation rate has already been reached for most of the parameter space \cite{Nguyen:2022zwb}, we will assume the above condition for our study. From eq.~(\ref{eq:capturerate}), we know that the annihilation rate is proportional to the local DM density, $\Gamma_\text{ann} \propto n_\chi$, and the total capture rate from eq.~(\ref{eq:ctotal}) is proportional to the number density of the celestial body, $\Gamma_\text{ann} \propto n_\star$, thus, the total annihilation rate is $\Gamma_\text{ann} \propto n_\chi n_\star$.

We consider the case in which DM annihilates into a long-lived ALP which weakly interacts with the matter such that it can escape from the celestial body and subsequently decay into a pair of SM particles. DM captured in celestial objects is expected to be annihilated with a small relative velocity, the mediator may be produced with the Lorentz boost factor, $\eta = m_\chi/m_a$, where $m_a$ is the mass of ALP. In order to calculate the sensitivities for the possible signals, we assume that the lifetime of the mediator, $\tau$, is sufficiently long such that the decay length exceeds the radius of the object $R$: 
\begin{eqnarray}
    L = \eta c \tau > R. 
\end{eqnarray}

The differential energy flux of gamma-rays/neutrinos arriving at Earth is given by
\cite{Leane:2017vag}
\begin{eqnarray}
    E^2 \frac{d\Phi}{dE} = \frac{\Gamma_\text{ann}}{4\pi D^2}\times E^2\frac{dN}{dE}\times \text{BR}(a\rightarrow \gamma\gamma/\nu\bar{\nu})\times P_\text{surv},\label{eq:flux}
\end{eqnarray}
where $D$ is the average distance from the target to the Earth. In this case, since the target is located around the galactic center, we assume $D\approx8$ kpc. Note also that our analysis is conservative as we do not consider gamma-ray or neutrino fluxes from a cascade decay of other particles.
%BR$(a\rightarrow \text{SM})$ is the branching ratio for the mediator into SM particles \textcolor{red}{which given by eq.~(\ref{eq:BRg}).}
%\textcolor{blue}{For simplicity, we are assuming the unity of the branching ratio and the analysis is then separated into 2 cases, i.e., the spectrum of DM annihilates into a mediator and then decays into a pair of either 100\% gamma-rays or 100\%  neutrinos.} 
The number distributions of the process $\chi\bar{\chi} \rightarrow aa \rightarrow 4\gamma$ can be described by a box-shaped spectrum \cite{Ibarra:2012dw} which is given by
%Their number distributions can be described by a box-shaped spectrum \cite{Ibarra:2012dw} which is given by
\begin{eqnarray}
    \frac{dN}{dE} = \frac{4}{\Delta E}\Theta(E - E_-)\Theta(E_+ - E),
\end{eqnarray}
where $\Theta$ is Heaviside step function, $\Delta E = E_+ - E_-$ is the width of the spectrum and $E_\pm = (m_\chi/2)\left(1\pm\sqrt{1-m_a^2/m_\chi^2}\right)$. The probability of the mediator, $a$, decaying outside the celestial object is
\begin{eqnarray}
    P_\text{surv} = e^{-R/\eta c \tau} - e^{-D/\eta c \tau}.
\end{eqnarray}

\begin{figure}[h]
\centering
    \includegraphics[height=6.5cm]{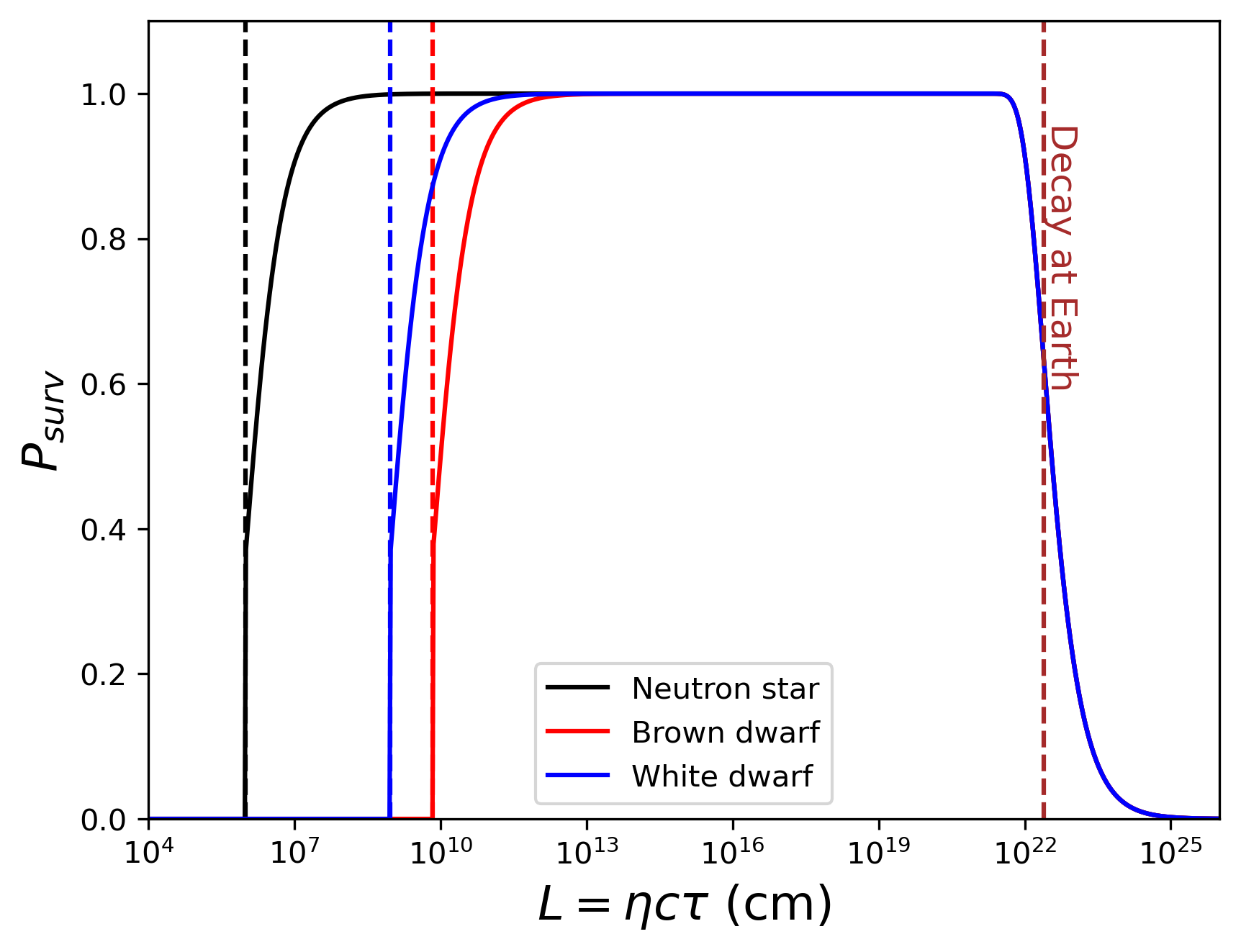}
    \caption{The plot shows the probability of the SM particle surviving from the NS (black), BD (red) and WD (blue) to the detector at the Earth. The object is assumed to be located at the center of the galaxy.}
    \label{fig:psurv}
\end{figure}
\begin{figure}[h]
\centering
    \includegraphics[height=6.5cm]{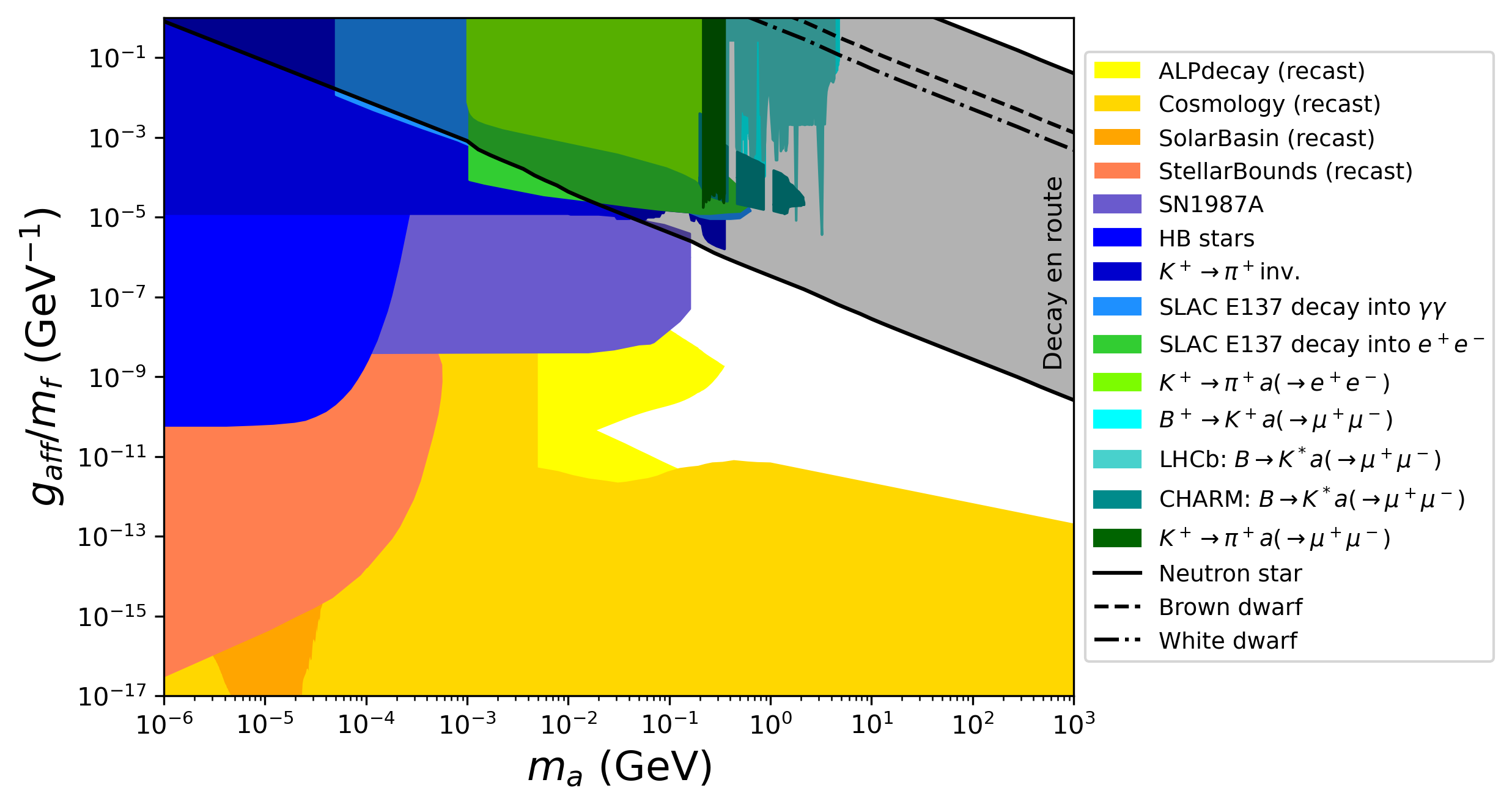}
    \caption{%The left panel shows the bounds on an effective ALP-photon coupling for an ALP decay between the Earth and the surface of NS (dash-dot line) BD (solid line) and WD (dash line). 
    The plot shows the upper bounds on the effective ALP-fermion coupling, $g_{aff}$, for ALP decay after escaping the surface of NS (solid line), BD (dashed line) and WD (dashed-dotted line). The lower bounds for ALP decay before reaching the earth is represented by the lower black solid line. The gray area between 2 solid lines shows the allowed region for the NS case.
    %, where in this case ALP decays directly into neutrinos.
    The constraints on $g_{a\gamma\gamma}$ are converted into $g_{aff}$ using eq.~(\ref{eq:relate}) with $B=0.1$. The other constraints on ALP-photon/SM coupling are discussed in the text.}
    \label{fig:gbound}
\end{figure}

An example of the survival probability as a function of the decay length $\eta c \tau$ is shown in figure \ref{fig:psurv}. The plot shows the survival probability of the SM particle produced by the mediator at the distance $\eta c \tau$ from the celestial objects. The probability is taken with the assumption that the SM particle does not reach the Earth if the mediator decays inside the star (dashed line), i.e., for $\eta c \tau < R$, the probability is assumed to be zero.
%for $\eta c \tau < R_\star$, the signal is assumed to be zero. The gamma-rays do not reach the Earth if they decay inside the star (red dash line) or decay behind the Earth (blue dash line).
%In this work, we will consider the maximally optimistic case where we take $P_\text{surv} = 1$ for both gamma-rays and neutrinos cases. 
Furthermore, we also calculate the bounds of %an effective ALP-photon coupling and 
an effective ALP-fermion coupling, $g_{aff}$, for ALPs that decay in the distance between the object's surface and the Earth using eq.~(\ref{eq:lifetime_nu}). 
%The bound on $g_{a\gamma\gamma}$ are calculated from the lifetime in equation \ref{eq:lifetime} 
%\textcolor{red}{We obtained the bound on $g_{a\gamma\gamma}$ by first we calculated $g_{aff}$ from the lifetime in equation \ref{eq:lifetime_nu} and convert to $g_{a\gamma\gamma}$ by the relation in equation \ref{eq:BR}}
%and the result is shown in the left panel of figure \ref{fig:gbound}. 
%We found that for the lower mass of ALP, the $g_{a\gamma\gamma}$ has already been excluded by the constraint from the helioscope CAST (CERN Axion Solar Telescope) \cite{CAST:2008ixs,CAST:2011rjr,CAST:2013bqn,CAST:2017uph}. Therefore for the ALP decaying into 100\% photons case, we will aim to study for ALP mass range higher than $10^{-9}$ GeV. 
%\textcolor{red}{We found that for the lower mass of ALP, the $g_{a\gamma\gamma}$ has already been excluded by various constraints \cite{AxionLimits}. Therefore, for the ALP decaying into %\textcolor{blue}{100\%} 
%photons case, we will aim to study the higher ALP masses. In the %\textcolor{blue}{100\% decaying into} neutrinos case, 
%\textcolor{red}{The} bound on $g_{aff}$ is calculated with the \textcolor{red}{decay width} from eq.~(\ref{eq:lifetime_nu}). 
The result is shown in the decay en route region in figure \ref{fig:gbound} where $m_f=m_\nu=0.1$ eV is assumed. In this work, we will consider the maximally optimistic case where we take $P_\text{surv} = 1$ when the decay width is within the range between the surface of the object and the distance to the surface of the Earth. 
We show the bounds from NS (solid line), BD (dashed line) and WD (dashed-dotted line) along with the constraints on the $g_{aff}$ (blue-green shaded). Additionally, we converted the ALP-photon coupling, $g_{a\gamma\gamma}$, to ALP-fermion coupling, $g_{aff}$, using eq.~(\ref{eq:relate}) with $B=0.1$ and displayed them in the orange-yellow shaded area. The details of the other constraints will be discussed in the following paragraphs.

%\textcolor{red}{We only show the bound from NS  and WD because NS cannot generate sufficient fluxes of neutrinos that are detectable by the experiments, which we will discuss shortly.}
%because they the only target that can generate sufficient fluxes of neutrinos that are detectable by the experiments, which we will discuss shortly.  

The recast constraints on $g_{a\gamma\gamma}$ shown in the orange-yellow shaded area of figure \ref{fig:gbound}, are adapted from ref.~\cite{AxionLimits} and summarized as follows. The ALP decay has included constraints from various sources, including diffuse gamma-ray \cite{Caputo:2022mah, Calore:2021hhn}, SN1987A-$\gamma$ (ALP decay) \cite{Jaeckel:2017tud, Hoof:2022xbe}, SN1987A-$\gamma,\nu$ (high mass ALPs) \cite{Lucente:2020whw, Caputo:2021rux, Caputo:2022mah}, MUSE telescope \cite{Todarello:2023hdk}, James Webb Space Telescope \cite{Janish:2023kvi}, VIMOS telescope \cite{Grin:2006aw}, Hubble Space Telescope \cite{Nakayama:2022jza,Carenza:2023qxh}, gamma-ray attenuation (ALP DM) \cite{Bernal:2022xyi}, XMM-Newton obsevations \cite{Foster:2021ngm}, INTEGRAL observations \cite{Calore:2022pks}, NuSTAR observations \cite{Perez:2016tcq,Ng:2019gch,Roach:2022lgo} and the heating of gas-rich dwarf galaxies Leo T \cite{Wadekar:2021qae}. The cosmology constraints include the constraints on ionisation fraction, extragalactic background light, X-rays \cite{Cadamuro:2011fd} and Big Bang Nucleosynthesis$+N_\text{eff}$ \cite{Depta:2020wmr}. The constraint on solar basin was taken from \cite{Beaufort:2023zuj}. The constraints on stellar bounds include the study of globular cluster \cite{Ayala:2014pea,Dolan:2022kul}, solar neutrinos \cite{Vinyoles:2015aba} and magnetic white dwarfs \cite{Dessert:2021bkv,Dessert:2022yqq}. %And the constraints on CAST Helioscopes were taken from \cite{CAST:2007jps,CAST:2017uph}.

The other constraints on $g_{aff}$ shown in the blue-green shaded area of figure \ref{fig:gbound} are adopted from the limit presented in ref.~\cite{Bharucha:2022lty}, where they concentrate on the case of strictly flavor-universal axion-fermion couplings. The constraints of the study of the supernova SN 1987A can provide strong constraints \cite{Jaeckel:2017tud} and recently updated in ref.~\cite{Hoof:2022xbe}. Considering the cooling of the horizontal branch (HB) stars provides the parameters space of ALPs \cite{Cadamuro:2011fd}. The exclusion bounds for the decay of $K^+ \rightarrow \pi^+ +$ inv. from NA62 \cite{NA62:2021zjw} are given in the form of branching ratio. The strong constraints come from the electron beam dump experiment SLAC E137 \cite{Bjorken:1988as}, ALPs can be produced by the bremsstrahlung, positrons annihilation, or Primakoff effect. $K^+ \rightarrow \pi^+ a(\rightarrow e^+e^-)$ from NA 48/2 \cite{NA482:2009pfe} provide C.L. 90\% upper limits. $B^+ \rightarrow K^+ a(\rightarrow \mu^+\mu^-)$ from LHCb \cite{LHCb:2016awg} provide C.L. 95\% upper limits on the branching ratio as a function of the lifetime of $a$. $B \rightarrow K^* a(\rightarrow \mu^+\mu^-)$ from LHCb \cite{LHCb:2015nkv} also provide the 95\% C.L. upper limit on the branching ratio as a function of the mass and lifetime of $a$. The limits from $B \rightarrow K^* a(\rightarrow \mu^+\mu^-)$ at fixed target come from CHARM \cite{Dobrich:2018jyi}. $K^+ \rightarrow \pi^+ a(\rightarrow \mu^+\mu^-)$ from NA 48/2 \cite{NA482:2016sfh} provide C.L. 90\% upper limits on the branching ratio as a function of lifetime of $a$. From the work of \cite{Bharucha:2022lty} where they have studied the model of the production of ALP-mediated DM in the thermal scenarios, our work (small $g_{a\chi\chi}$ and $g_{aff}$) corresponds to the case that DM is produced by freeze-in and the ALP may or may not be in equilibrium with the SM. Therefore, the viable parameter space for the freeze-in scenario is not ruled out by our study.
%%% This is 

In order to set the limits on the ALP coupling in our model, we use the galactic center measured fluxes from various observations. The gamma-ray fluxes are taken from Fermi and H.E.S.S. and the details of the data analysis are given in \cite{Malyshev:2015hqa}. Fermi data is used for the model fluxes with energy less than 100 GeV \cite{Fermi-LAT:2015bhf}, and we use H.E.S.S. \cite{HESS:2016pst} to set limits around TeV scale. The high energy neutrino fluxes were taken from IceCube \cite{IceCube:2017trr} and ANTARES \cite{ANTARES:2016mwq} where they have the upper limits on the muon neutrinos flux from the direction of the galactic center ($-40^\circ< l< 40^\circ$ and $-3^\circ< b< 3^\circ$ in Galactic Coordinate) in the energy range between 1 TeV and 1 PeV. We set the limits by requiring that the maximum value of the DM fluxes must not exceed the measurable fluxes from the observation. This can be done by determining %the ALP coupling 
$g_{aff}$ for each energy bin $(E \approx m_\chi)$ which gives the highest possible scattering cross-section from eq.~(\ref{eq:xsec}) providing the energy flux in eq.~(\ref{eq:flux}) below the measured flux.

\begin{figure}[h]
\centering
    \includegraphics[height=6.5cm]{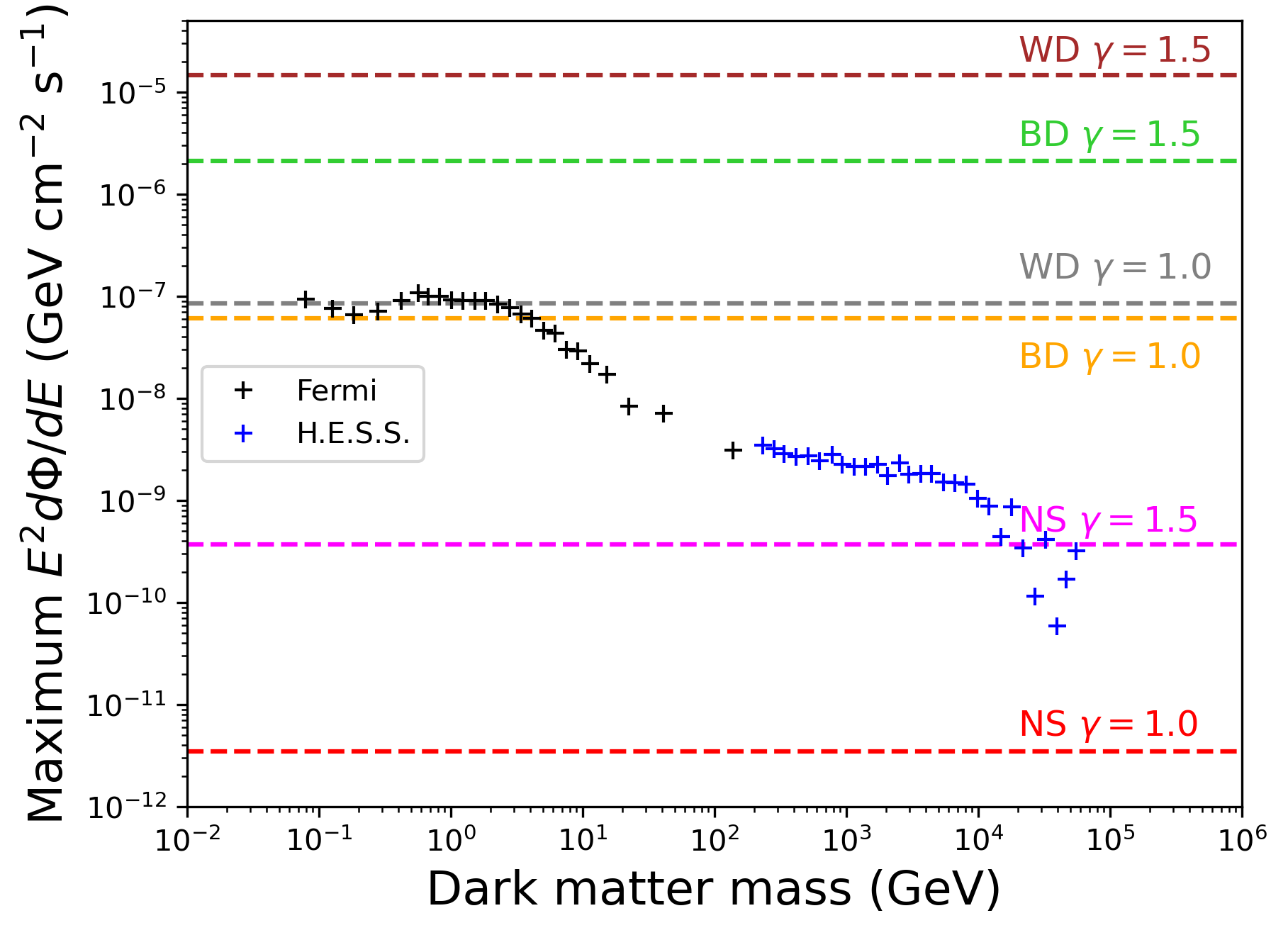}
    \caption{Maximum differential fluxes of gamma-ray of the NSs, BDs and WDs from the galactic center. The gamma-ray fluxes from the galactic center observed by Fermi and H.E.S.S. are shown in black and blue, respectively. }
    \label{fig:limit}
\end{figure}

To generate the gamma-ray flux
%energy spectra 
from eq.~(\ref{eq:flux}), we calculate the total DM capture rate from eq.~(\ref{eq:xsec}) by using  \texttt{Asteria} \cite{Leane:2023woh}. We use the number density of NSs from eq.~(\ref{eq:NS}) and the number density of BDs from eq.~(\ref{eq:BD}). We integrate both densities over the spherical shell between radius $r_1=0.1$ pc and $r_2=100$ pc. Choosing $r_1 = 0.1$ pc to avoid poor modeling of the DM cusp-like profile at the galactic center. The outer radius $r_2 = 100$ pc is chosen due to the fact that NSs/BDs number density falls rapidly outside this region. In WDs case, we choose $r_1=10^{-4}$ pc and $r_2=1.5$ because the number density of WD from eq.~(\ref{eq:WD}) is only valid in this area.
%We assume that the ALP decays into 100\% photons and $P_\text{surv}=1$ for simplicity. 
The breaching ratio $\text{BR}(a\rightarrow\gamma\gamma)$ was taking from eq.~(\ref{eq:BRg}) with $B=0.1$ and we assume $P_\text{surv}=1$ for simplicity.
We show the largest possible signal for different choices of density slope index $\gamma=1,1.5$ in the NFW DM profile in figure \ref{fig:limit} along with the measured gamma-ray fluxes from Fermi (black) and H.E.S.S. (blue). For NSs, one could expect that only the gamma-rays from the NFW DM profile with $\gamma=1.5$ are detectable by H.E.S.S. data for a short range of DM masses, $ m_\chi \gtrsim 2\times10^4$ GeV.
%would come from the NFW DM profile with $\gamma=1.5$ for the mass over 100 GeV which corresponds to H.E.S.S. data. 
%For BDs, both gamma-rays from slope index $\gamma=1, 1.5$ are higher than the measured fluxes from Fermi and H.E.S.S., which would give us strong constraints to the ALP coupling. 
For BDs, the gamma-rays from slope index $\gamma= 1.5$ are higher than the measured fluxes from Fermi and H.E.S.S., and with $\gamma = 1$, the gamma-ray would come from a full range of H.E.S.S. and a short range of Fermi, $ m_\chi \gtrsim 3$ GeV.
For WDs, both gamma-rays from slope index $\gamma=1, 1.5$ are higher than the measured fluxes from Fermi and H.E.S.S., which would give us strong constraints to the ALP-fermion coupling.

\begin{figure}[h]
\centering
    \includegraphics[height=6.5cm]{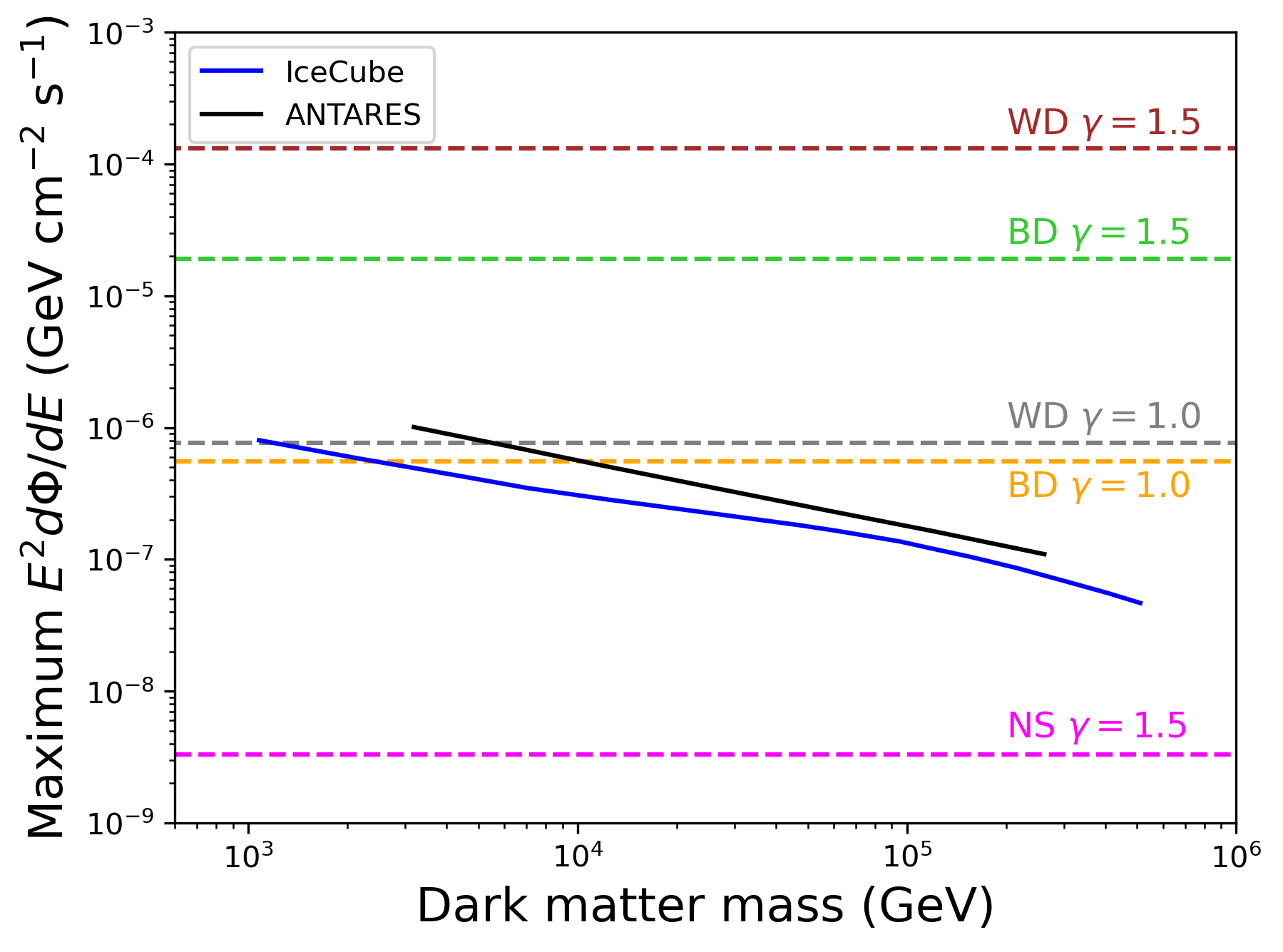}
    \caption{Maximum differential fluxes of neutrino from the NSs, BDs and WDs in the galactic center. The neutrino fluxes from the galactic center observed by IceCube and ANTARES are shown in blue and black, respectively.}
    \label{fig:limit_nu}
\end{figure}

To generate the neutrino  %spectra
flux, we follow the same procedure as the gamma-ray %spectra 
flux. The largest possible fluxes for $\gamma=1, 1.5$ are shown in figure \ref{fig:limit_nu} along with the measured neutrino fluxes from IceCube (blue) and ANTARES (black). % We find that the maximum flux from the NS is lower than the observation fluxes by approximately 2 orders of magnitude.
We find that the maximum flux from the NS appears to be lower than the observation fluxes by approximately 2 orders of magnitude, therefore, we will not consider NS in the neutrino case. Note that our neutrino fluxes seem lower than those of \cite{Nguyen:2022zwb, Bose:2024wsh}. The maximum flux from BDs with $\gamma = 1$ is lower than the short range of the measured flux ($m_\chi\lesssim3\times10^3$ GeV for Icecube and $m_\chi\lesssim 10^4$ GeV for ANTARES), where BDs with $\gamma=1.5$ can be observed by both observations. The maximum fluxes from WDs with $\gamma=1.5$ are higher than the measured flux and WDs with $\gamma = 1$ are slightly lower than the measured flux ($m_\chi\lesssim 1.2\times10^3$ GeV for Icecube and $m_\chi\lesssim 6\times10^3$ GeV for ANTARES).  As a result, we will only study BDs and WDs as the potential sources of neutrino fluxes.

%In addition, we studied the possibility that ALP decays into two channels (neutrinos and photons) and we discovered that the neutrino will dominate the decaying and suppress the gamma-ray flux. Consequently, we will study the gamma-rays and neutrinos independently.

\section{Results}

\begin{figure}[h]
\centering
\begin{subfigure}{0.87\textwidth}
    \includegraphics[width=\textwidth]{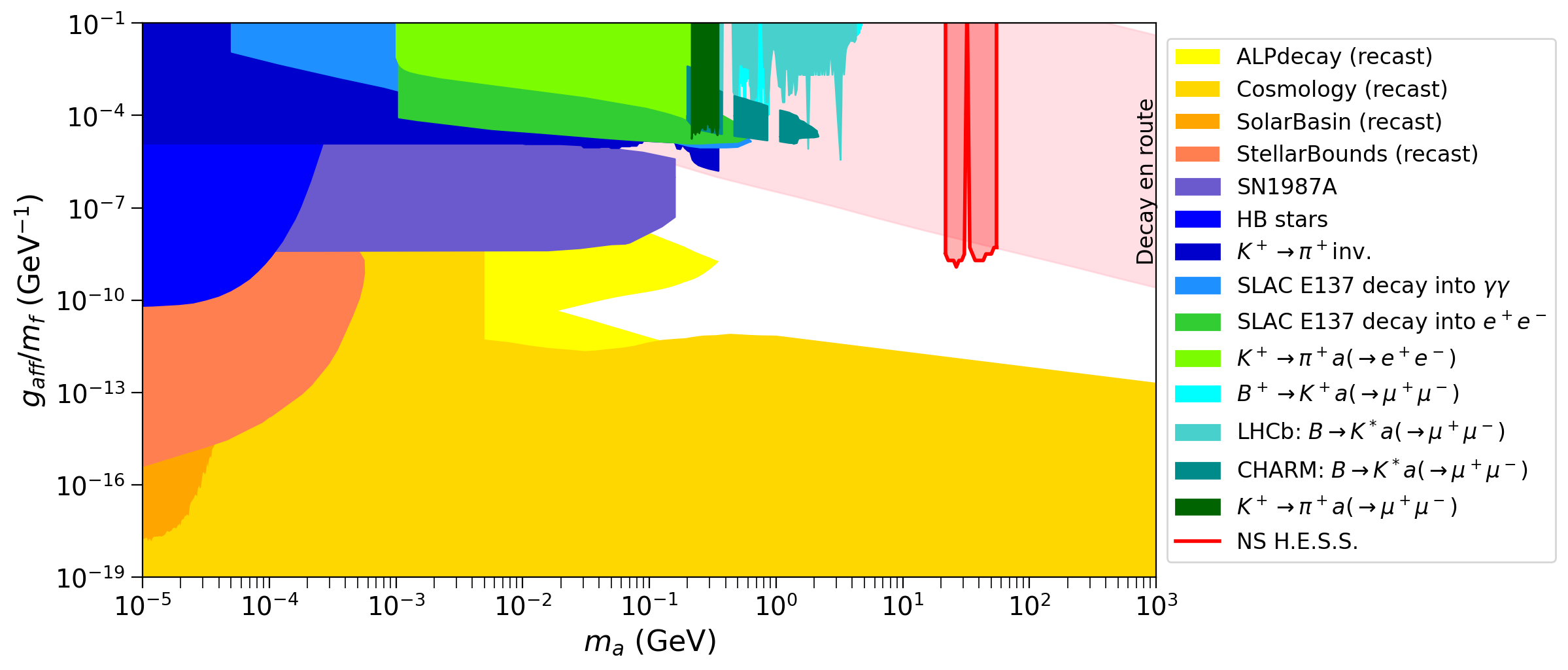}
    \caption{\footnotesize Neutron star}
    \label{fig:gammaNS}
\end{subfigure}

\begin{subfigure}{0.87\textwidth}
    \includegraphics[width=\textwidth]{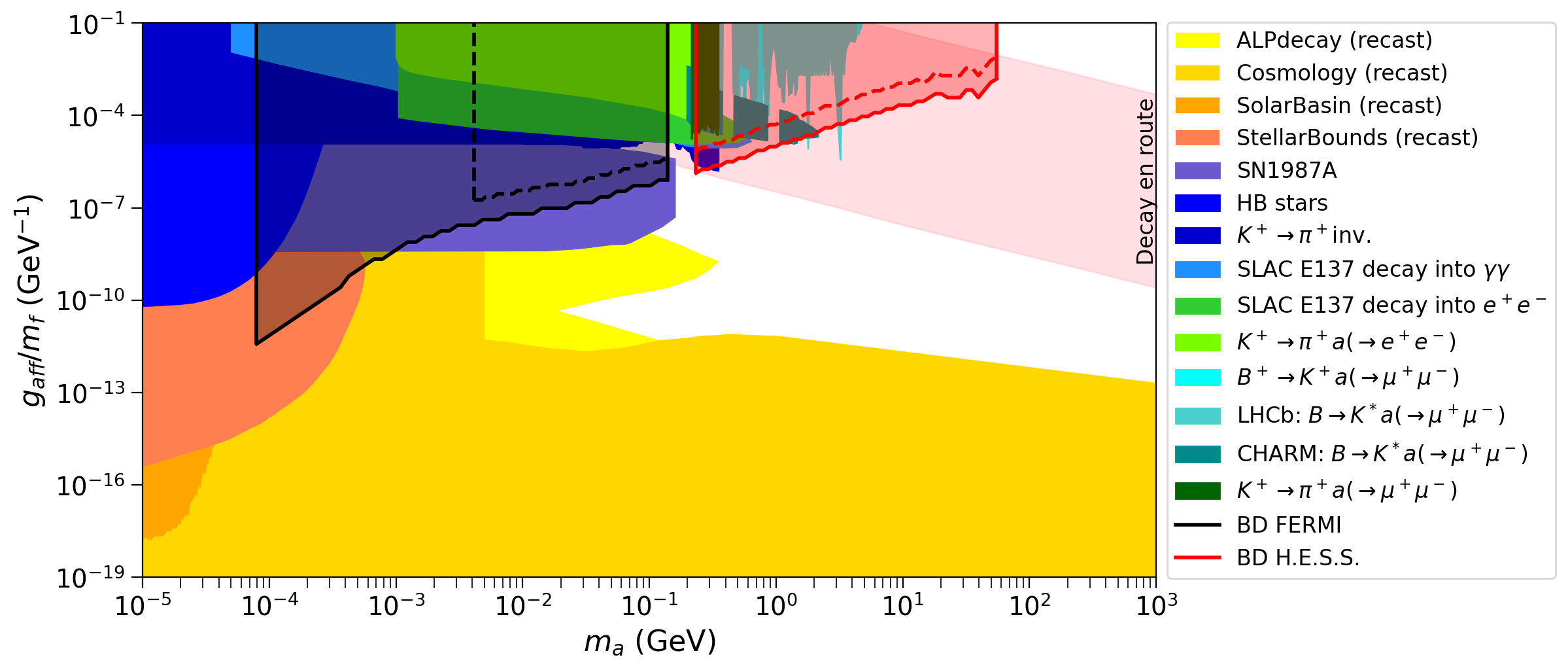}
    \caption{\footnotesize Brown dwarf}
    \label{fig:gammaBD}
\end{subfigure}

\begin{subfigure}{0.87\textwidth}
    \includegraphics[width=\textwidth]{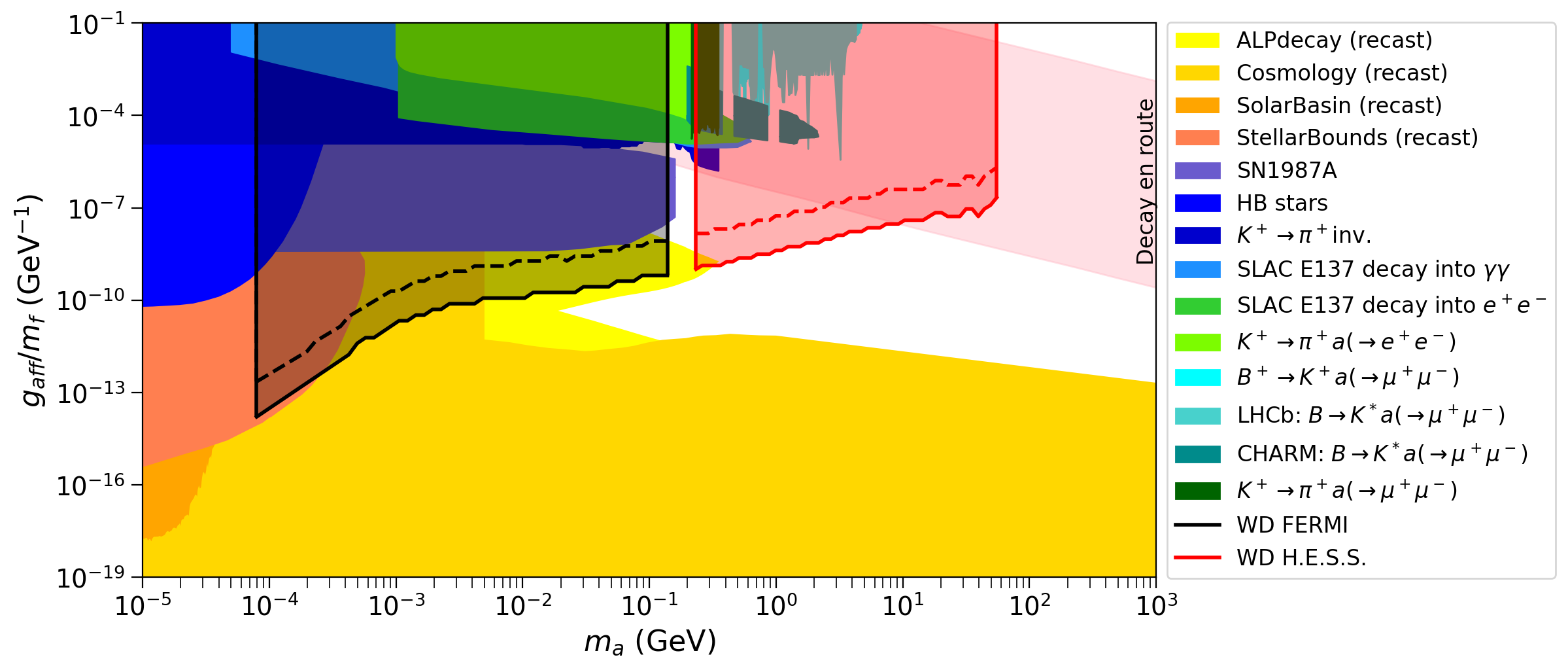}
    \caption{\footnotesize White dwarf}
    \label{fig:gammaWD}
\end{subfigure}
        
\caption{The constraints plot on the ALP-fermion coupling, $g_{aff}$ from gamma-ray observations as a function of $m_a$ where we fixed $g_{a\chi\chi} =10^{-3}$, $B=0.1$ and $\eta = m_\chi/m_a = 10^3$. The DM profile with $\gamma = 1$ and $\gamma = 1.5$ were shown in dashed and solid lines. The results on H.E.S.S. and Fermi are shown in black and red, respectively.}
\label{fig:results1}
\end{figure}

% CP attempt

In the following analysis, the set of model parameters ($g_{aff}$, $B$, $m_a$, $\eta$, $g_{a\chi\chi}$) has been chosen to describe the ALP-fermion coupling, the branching ratio from ALP to gamma-rays, the ALP mass, the boosting factor, and the ALP-DM coupling respectively. We utilize \texttt{Asteria} to calculate the capturing rate and then generate both gamma-rays and neutrino fluxes to compare with constraints from Fermi, H.E.S.S., IceCube, and ANTARES. 
%The gamma-rays were produced by the production of DMs that annihilated inside the object, $\chi\chi\rightarrow a$, $a\rightarrow\gamma\gamma$. The limits were placed on $E^2 d\Phi/dE$ with the gamma-ray observatories (Fermi and H.E.S.S.). Our calculation on gamma-ray flux has been described in the previous section and the other parameters in the plot are \textcolor{red}{$g_{a\chi\chi}=1\times10^{-3}$ and $\eta = m_\chi/m_a = 10^4$. 
%The bounds from our works are presented along with other constraints on $g_{aff}$. Since the $g_{aff}$ is related to $g_{a\gamma\gamma}$ via the definition of the branching ratio in eq.~(\ref{eq:relate}), we have transformed the constraints on $g_{a\gamma\gamma}$ to $g_{aff}$ and displayed them along with the results. The Fermi upper limits are shown in black and the H.E.S.S. upper limits are shown in red. We display the results for DM profile with $\gamma=1,1.5$ in dash and dot lines respectively.}
%We show the results of BDs for $\gamma =1$ in the dash line and $\gamma=1.5$ in the solid line. 
%Since we concentrate on the simplest case where $C_f$ is universal, therefore the result we present is in the form of $g_{aff}/m_f = C_f/f_a$, where $m_f$ is the mass of the fermion. In the gamma-rays final state, we use proton and neutron mass for calculating the scattering cross-section, therefore $m_f = m_n \approx m_p$. 
Figure \ref{fig:results1} shows our constraints on $g_{aff}$ where $B = 0.1$, $g_{a\chi\chi} = 10^{-3}$ and $\eta = 10^3$. 
%We also highlight the region forbidden by the perturbative limit, $g_{a\chi\chi}\geq 4\pi^2$ in the black hatched area.
For NS (figure \ref{fig:gammaNS}), we show only $\gamma=1.5$ since the gamma-ray flux is detectable only at H.E.S.S. for $ m_\chi \gtrsim 2\times10^3$ GeV. 
%Our results show that the limits in this work are significantly stronger than previously existing constraints. 
%Our results show that the limits in this work are significantly stronger than previous ALP-SM coupling constraints. 
For lower masses, both BDs and WDs can provide stronger bounds than the previous ALP-fermion coupling constraints where WDs give the most strongest bounds. However, these constraints are still being ruled out by the recasting constraints from $g_{a\gamma\gamma}$.
For higher masses, our results indicate that the limits in this work are stronger than the other constraints, especially NSs and WDs. Moreover, under the universal ALP-fermion coupling assumption, the region where ALP decays en route (figure \ref{fig:gbound}) can be constrained up to $m_a\sim 60$ GeV.
%by BDs for $m_a \gtrsim 0.5$ GeV.
%The results show that WDs provide stronger bounds than NSs and BDs since WDs require a lower cross-section to capture the DM (figure \ref{fig:Cmax})
%The results of BDs were shown in \ref{fig:gammaBD}. The results of WDs were shown in \ref{fig:gammaWD}.
%For higher masses, NSs provide stronger bounds than BDs since NSs require a lower cross-section to capture the DM (figure \ref{fig:Cmax}). 

\begin{figure}[h]
\centering
\begin{subfigure}{0.87\textwidth}
    \includegraphics[width=\textwidth]{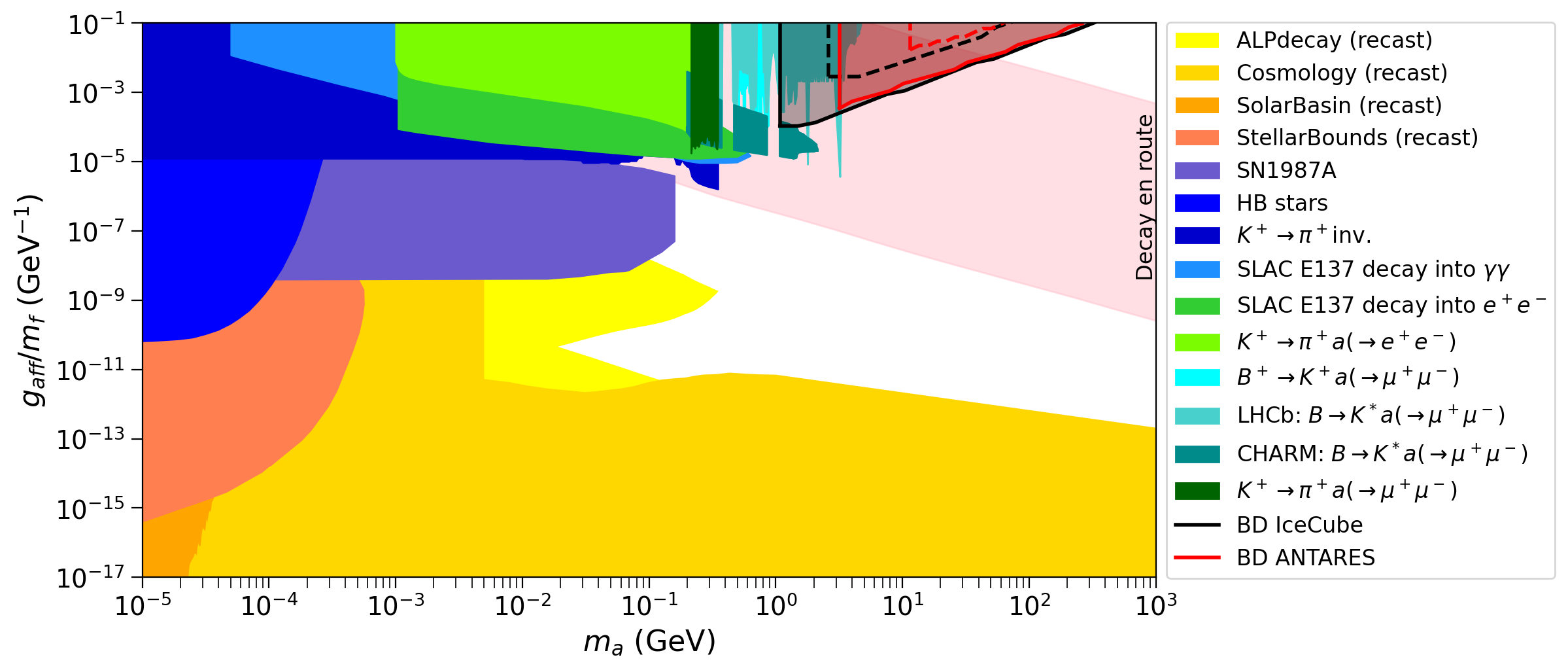}
    \caption{\footnotesize Brown dwarf.}
    \label{fig:nuBD}
\end{subfigure}

\begin{subfigure}{0.87\textwidth}
    \includegraphics[width=\textwidth]{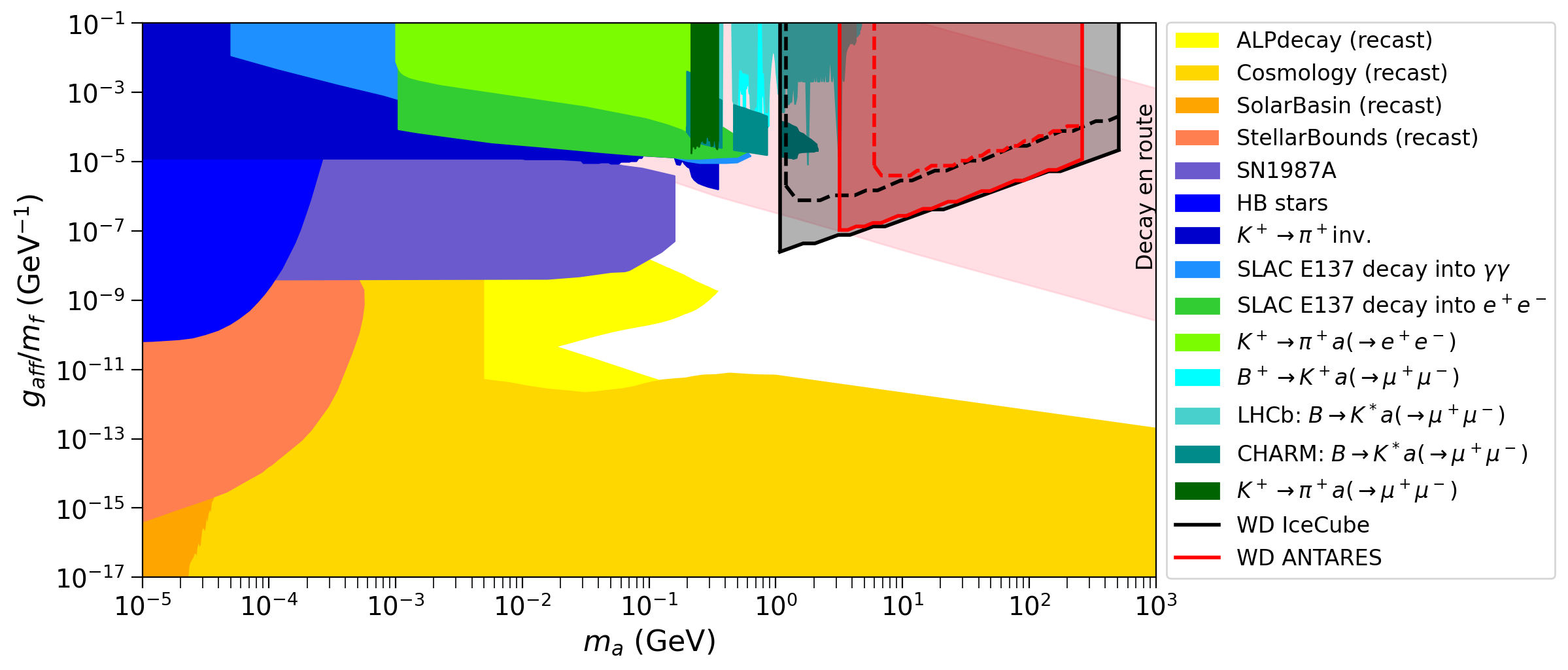}
    \caption{\footnotesize White dwarf.}
    \label{fig:nuWD}
\end{subfigure}
        
\caption{The constraints plot from the neutrino observations on the ALP-fermion coupling, $g_{aff}$ as a function of $m_a$ where we fixed $g_{a\chi\chi} = 10^{-3}$, $B=0.1$ and $\eta = m_\chi/m_a = 10^3$. The NFW DM profiles with $\gamma=1, 1.5$ are represented in dashed and solid lines, respectively. The results on IceCube and ANTARES are shown in black and red, respectively. }
\label{fig:results2}

\end{figure}

Figure \ref{fig:results2} shows the upper limits on the ALP-fermion coupling from the neutrino observations (IceCube and ANTARES). The neutrino fluxes were generated by focusing on BD and WD as the targets as discussed in the earlier section. The bounds are weaker than the gamma-ray case but still stronger than existing constraints, especially in the higher ALP mass region. Under the universal ALP-fermion coupling assumption, the region where ALP decays en route can be constrained up to $m_a\sim 60$ GeV for BDs and up to $m_a\sim 500$ GeV for WDs.

%\textcolor{red}{for BDs can be constrained to $m_a \gtrsim 1.5$ GeV and for WDs can be constrained to $m_a \gtrsim 10$ GeV.}

\begin{figure}[h]
\centering
    \includegraphics[width=0.87\textwidth]{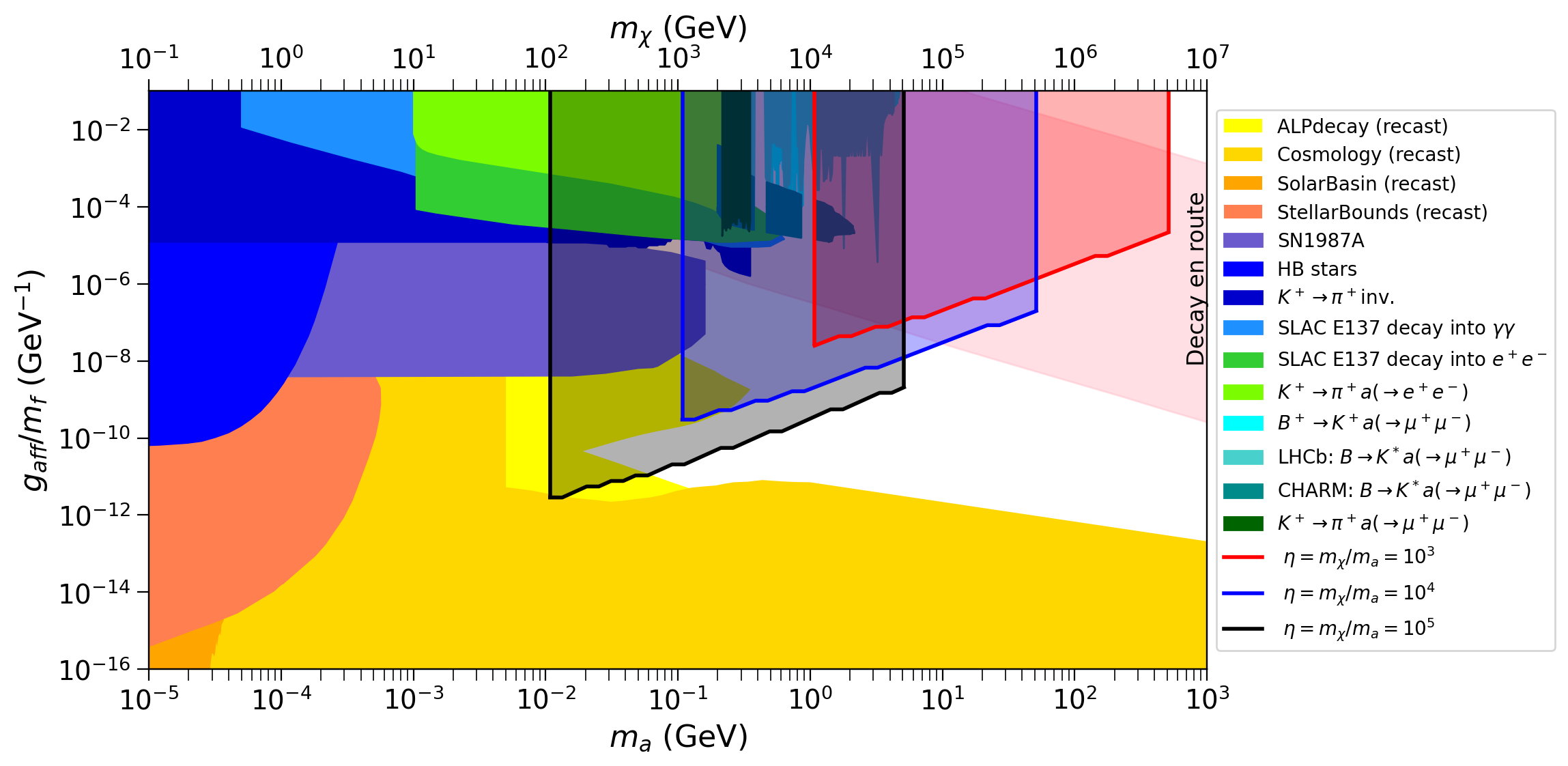}
    \caption{The constraints plot of the ALP-fermion coupling on the IceCube experiment where we fixed the $g_{a\chi\chi} = 10^{-3}$ and $B=0.1$. We only focused on the WDs as the only sources. We show the results for the mass $\eta = m_\chi/m_a = 10^3,\ 10^4$ and $10^5$ in red, blue and black, respectively.}
    \label{fig:changeR}
\end{figure}

We also show the results when we vary the ratio of the mass of DM and ALP in figure \ref{fig:changeR}. The constraints on $g_{aff}$ using WDs as the targets are calculated with $g_{a\chi\chi} = 10^{-3}$ and the mass ratios $\eta = m_\chi/m_a = 10^3,\ 10^4$ and $10^5$ are shown in red, blue and black, respectively. The results show that the decreasing of the mass ratio gives stronger bounds in the lower masses region of ALP.

\begin{figure}[h]
\centering
\begin{subfigure}{0.87\textwidth}
    \includegraphics[width=\textwidth]{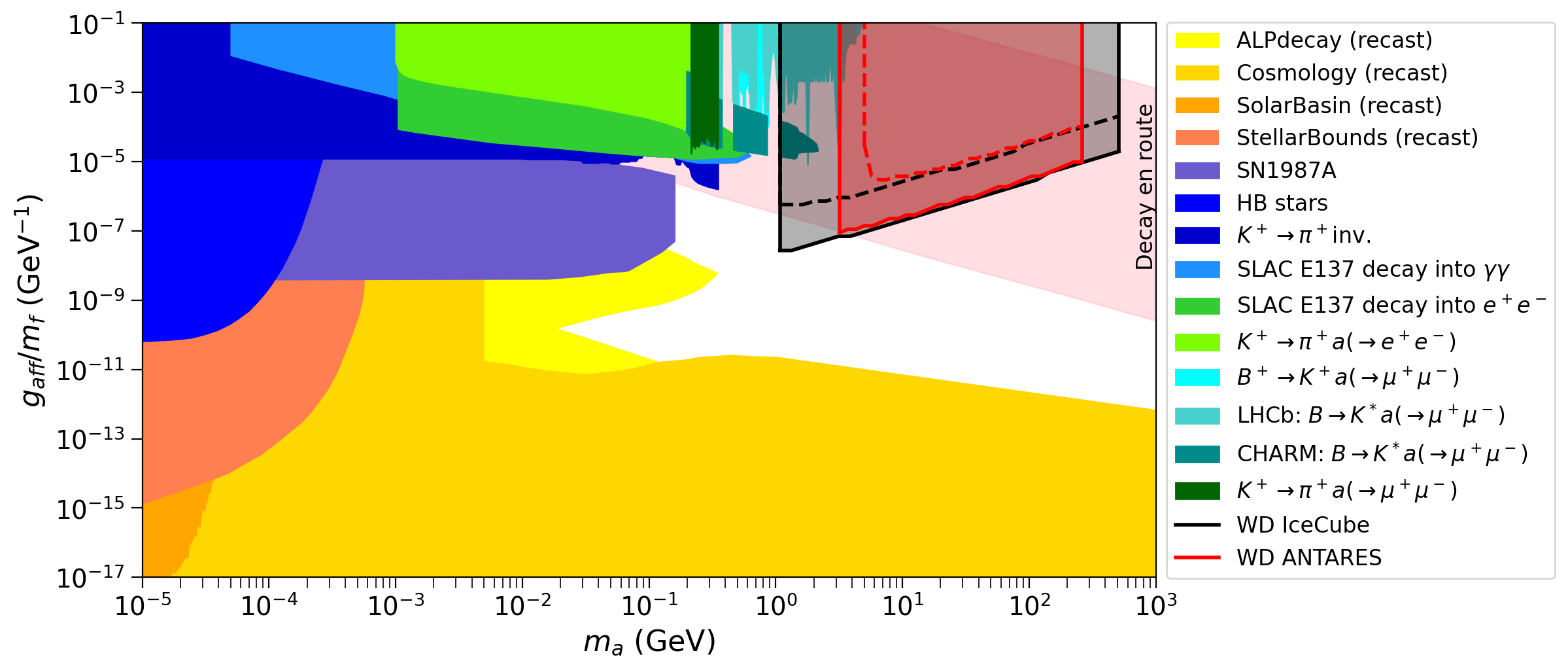}
    \caption{\footnotesize $B=0.01$}
    \label{fig:B001}
\end{subfigure}

\begin{subfigure}{0.87\textwidth}
    \includegraphics[width=\textwidth]{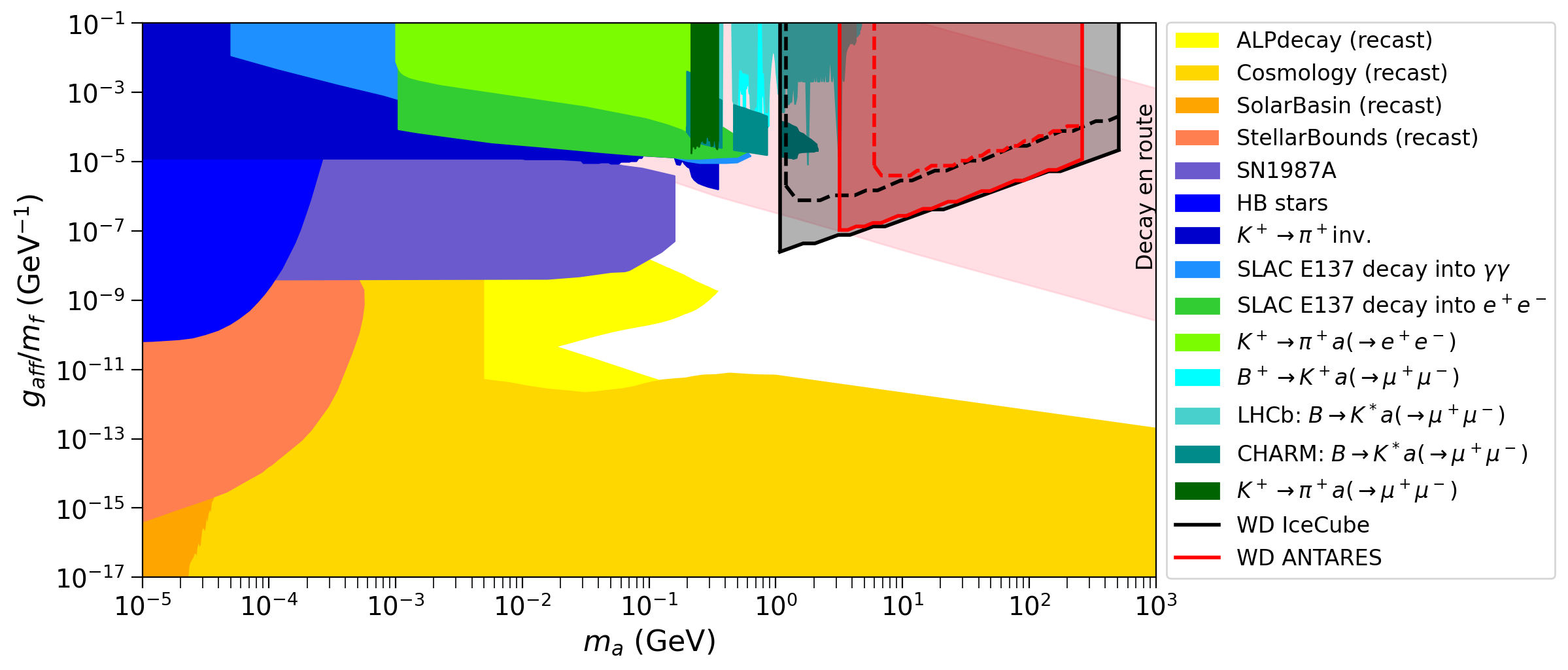}
    \caption{\footnotesize $B=0.1$}
    \label{fig:B01}
\end{subfigure}

\begin{subfigure}{0.87\textwidth}
    \includegraphics[width=\textwidth]{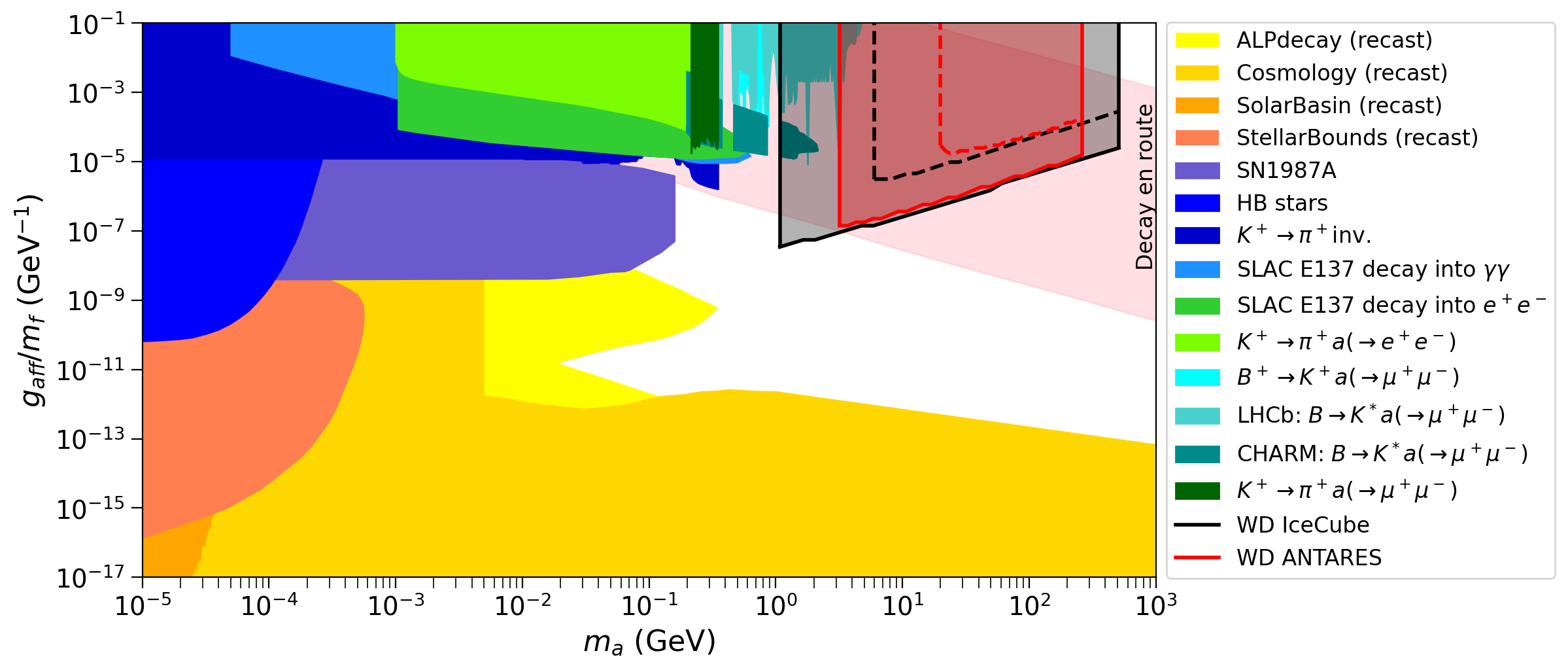}
    \caption{\footnotesize $B=0.5$}
    \label{fig:B05}
\end{subfigure}
        
\caption{The constraints plot of $g_{aff}$ from the neutrino observations by using WDs as the target where we fixed the $g_{a\chi\chi} = 1\times10^{-3}$ and $\eta = m_\chi/m_a = 10^3$. The plots are displayed with different values of the parameter $B=0.01,\ 0.1$ and $0.5$. The dashed and solid lines represent the NFW DM profiles with $\gamma=1,\ 1.5$, respectively.}
\label{fig:recast}

\end{figure}

In figure \ref{fig:recast}, we present the results obtained by varying the parameter $B$ in eq.~(\ref{eq:BRg}), where we display the limits of $g_{aff}$ versus ALP mass. The plots show the results from neutrino observations using WDs as the target
where %with $B=0.01,\ 0.1,\ 1$.
%We have recast the results from neutrino observations by using WDs as the target and the constraints on ALP-SM coupling to the ALP-photon coupling and displayed in the plot. 
the different choices of parameter $B$ are applied. The result indicates that increasing the parameter $B$ leads to elevated gamma-ray fluxes, which, on the other hand, also simultaneously decreases neutrino fluxes. Consequently, this leads to weaker bounds on $g_{aff}$ from the neutrino observations especially in the case where $\gamma = 1$.

%\newpage

\newpage

\section{Conclusion}

In this work, we have studied the constraints of the ALP-mediated DM models from celestial objects such as neutron stars (NSs), brown dwarfs (BDs) and white dwarfs (WDs). The DM can be captured in celestial objects where the DM accumulation process is controlled by the dark matter multiscatter capture mechanism. The annihilation of DM inside the objects produces the long-lived ALPs which will decay into the SM particles outside the objects. We study the case that ALP decays into gamma-rays and neutrinos %separately
and compare the fluxes with data from observations.
For gamma-ray fluxes, we present our results on $g_{aff}/m_f$ in figure \ref{fig:results1}. The results show the strong bounds that can rule out a significant portion of the parameter space up to $\sim\mathcal{O}(10)$ GeV mass range. We point out that by using WDs as targets the model provides a stronger bound than NSs and BDs %BD ones 
due to the lower fluxes required to match with the H.E.S.S. data. For Fermi data with the lower DM mass range, only BDs and WDs are detectable via the gamma-ray channel. %\textcolor{red}{?In the condition that ALP decay en route towards the Earth, the only target that can be constrained is BD with $m_a \gtrsim 0.5$ GeV.}
%We also present the upper bounds for different mass ratios in figure \ref{fig:changeR} and show that our limits are generally stronger than existing constraints across the mass range of ALP.
The results for neutrino fluxes are present in figure \ref{fig:results2}. We found that the %only 
possible targets that can produce sufficient neutrino flux are BDs and WDs. With data from IceCube and ANTARES, we can probe our model with the higher mass range. Similar to the gamma-ray channel, the bounds from neutrino fluxes are also generally stronger than the previous constraints on ALP. Provided that ALP decays en route towards the Earth, the allowed parameter space for this model is strongly constrained up to $\sim\mathcal{O}(100)$ GeV for BDs and WDs. This emphasizes the importance of the next-generation probes of the neutrino fluxes at higher energies such as ARIANNA, ARIA, IceCube-Gen2 \cite{IceCube-Gen2:2020qha,Ishihara:2019aao}, KM3Net \cite{KM3Net:2016zxf}, ANITA-IV \cite{ANITA:2019wyx}, PUEO \cite{PUEO:2020bnn}, RNO-G \cite{RNO-G:2020rmc} and Auger \cite{PierreAuger:2019ens}, as well as gamma-ray fluxes from Cherenkov Telescope Array (CTA) \cite{CTAConsortium:2013ofs,CTAConsortium:2017dvg}.

We %also 
present the upper bounds for boost factor, $\eta$, in figure \ref{fig:changeR} and show that our limits are generally stronger than existing constraints across the mass range of ALP. We also present the limits for different fractions of the branching ratios in figure \ref{fig:recast}. Our findings indicate that an increase of the gamma-ray flux will reduce the flux of the neutrinos which weakens the limits from neutrino observations.

For future works, we are interested in exploring the phenomenological consequences of the possibility that ALP decays inside celestial objects providing additional energy to their internal processes. We also would like to investigate the effects of ALP-mediated interactions on the nuclear equation of state (EoS) and neutron star stability.

\section*{Acknowledgement}
TK and CP have been supported by the National Astronomical Research Institute of Thailand (NARIT). This research project is supported by National Research Council of Thailand (NRCT) : (Contact No. N41A670401).  CP is supported by Fundamental Fund 2566 of Khon Kaen University and Research Grant for New Scholar, Office of the Permanent Secretary, Ministry of Higher Education, Science, Research and Innovation under contract no. RGNS64-043. We thank Areef Waeming and Daris Samart for their helpful comments and discussions. We also thank the referees for their constructive comments and suggestions.

\appendix
\section{Appendices} 
\subsection{ALP interaction length inside the object} \label{sect:app1}
Consider the dominant scattering cross-section of ALP-neutron via s-channel which is given by
\begin{eqnarray}
    \sigma_{an} = g_{ann}^4\frac{\pi^4}{8 m_n} \frac{E_i + m_n}{(2E_i + m_n)^2},
\end{eqnarray}
where $g_{ann}$ is ALP-nucleon coupling, $m_n$ is the nucleon mass and $E_i$ is the energy of incoming ALP. The interaction length can be approximated by
\begin{eqnarray}
    l_a = \frac{1}{n\times\sigma_{an}}    
\end{eqnarray}
where $n = \rho_\star/m_n$ is the number density of nucleons in the star. We have assumed $g_{ann}=10^{-5}$ GeV, $E_i = 10^3$ GeV and we found that the interaction length in the NS is $1.98\times10^6$ km, in the BD is $2.51\times10^{19}$ km and in the WD is $4.69\times10^{15}$ km. This means the ALP would be able to escape the star before being trapped.

\bibliographystyle{jhep}
\bibliography{ref}

\end{document}